%% file: main_usenix.tex
\documentclass[letterpaper,twocolumn,10pt]{article}
\usepackage{usenix}

\usepackage{tikz}
\usepackage{amsmath}

\usepackage{booktabs}
\usepackage{array}
\usepackage{multirow}
\usepackage{tabularx}
\usepackage{array}
\usepackage{hyperref}
\usepackage{footnotehyper}
\makesavenoteenv{table}

\usepackage{enumitem}
\usepackage{booktabs}
\usepackage{wrapfig}
\usepackage{listings}
\usepackage[capitalize]{cleveref}
\usepackage{listings}
\usepackage{xurl}
\usepackage{seqsplit}

\usepackage[most]{tcolorbox}
\usepackage{listings}
\usepackage{xcolor}
\usepackage{caption}
\usepackage{graphicx}
\usepackage[section]{placeins}
\usepackage{multicol}
\newsavebox{\trustedapibox}

\definecolor{codebackground}{HTML}{303643}
\definecolor{codeforeground}{HTML}{F2F4F8}
\definecolor{codetitle}{HTML}{D8DEE9}
\definecolor{codeborder}{HTML}{D8DEE9}
\usepackage[most]{tcolorbox}
\usepackage{tikz}
\usetikzlibrary{arrows.meta, positioning, fit, calc}

\usepackage{xspace}
\newcommand{\sys}{Baya\xspace}
\newcommand{\reactApiWeb}{$\text{ReAct}_{\text{web + API}}$\xspace}
\newcommand{\reactWeb}{$\text{ReAct}_{\text{web}}$\xspace}
\newcommand{\sysNew}{$\text{Baya}_{\text{Code}}$\xspace}
\newcommand{\sysOld}{$\text{Baya}_{\text{JSON}}$\xspace}
\newcommand{\heading}[1]{\vspace{0.03in}\noindent\textbf{#1}}

\newcommand{\introrev}[1]{#1}
\newcommand{\revision}[1]{#1}

\newtcblisting{appendixcode}[1]{
    enhanced,
    breakable,
    listing only,
    title={#1},
    title after break={#1\space(Continued)},
    colback=white,
    colframe=black,
    colbacktitle=black,
    coltitle=white,
    coltext=black,
    fonttitle=\bfseries\small,
    boxrule=0.8pt,
    width=\dimexpr\linewidth-6pt\relax,
    center,
    arc=2mm,
    outer arc=2mm,
    left=1.5mm,
    right=1.5mm,
    top=1mm,
    bottom=1mm,
    listing options={
        language=Python,
        basicstyle=\ttfamily\small\color{black},
        keywordstyle=\bfseries\color{black},
        identifierstyle=\color{black},
        commentstyle=\color{black!65},
        stringstyle=\color{black},
        breaklines=true,
        breakatwhitespace=false,
        columns=fullflexible,
        keepspaces=true,
        showstringspaces=false,
        emptylines=0,
        aboveskip=0pt,
        belowskip=0pt
    }
}
\newtcblisting{appendixcodewide}[1]{
    enhanced,
    listing only,
    title={#1},
    colback=white,
    colframe=black,
    colbacktitle=black,
    coltitle=white,
    fonttitle=\bfseries\small,
    boxrule=0.8pt,
    arc=2mm,
    left=1.5mm,
    right=1.5mm,
    top=1mm,
    bottom=1mm,
    listing options={
        language=Python,
        basicstyle=\ttfamily\small,
        breaklines=true,
        breakatwhitespace=false,
        columns=fullflexible,
        keepspaces=true,
        showstringspaces=false
    }
}

\newtcblisting{toycode}[1]{
    enhanced,
    listing only,
    title={#1},
    colback=white,
    colframe=black!65,
    colbacktitle=black!10,
    coltitle=black,
    fonttitle=\bfseries\footnotesize,
    boxrule=0.5pt,
    arc=1mm,
    left=1.2mm,
    right=1.2mm,
    top=0.8mm,
    bottom=0.8mm,
    before skip=0pt,
    after skip=0pt,
    listing options={
        language=Python,
        basicstyle=\ttfamily\fontsize{6.5pt}{7.1pt}\color{black},
        keywordstyle=\bfseries\color{black},
        commentstyle=\color{black!65},
        stringstyle=\color{black},
        breaklines=true,
        columns=fullflexible,
        keepspaces=true,
        showstringspaces=false,
        emptylines=1,
        aboveskip=0pt,
        belowskip=0pt
    }
}

\usepackage{tikz}
\usetikzlibrary{arrows.meta}

\newtcblisting{toylog}[1]{
    enhanced,
    listing only,
    listing options={
        basicstyle=\ttfamily\scriptsize\color{black},
        breaklines=true,
        columns=fullflexible,
        keepspaces=true,
        showstringspaces=false,
        mathescape=false,
        emptylines=1,
        aboveskip=0pt,
        belowskip=0pt
    },
    title={#1},
    colback=white,
    colframe=black!65,
    colbacktitle=black!10,
    coltitle=black,
    fonttitle=\bfseries\footnotesize,
    boxrule=0.5pt,
    arc=1mm,
    left=1.2mm,
    right=1.2mm,
    top=0.8mm,
    bottom=0.8mm,
    before skip=0pt,
    after skip=0pt,
}

\newcommand{\compilerbot}{%
    \raisebox{-0.25\height}{%
        \includegraphics[
            height=5.5mm
        ]{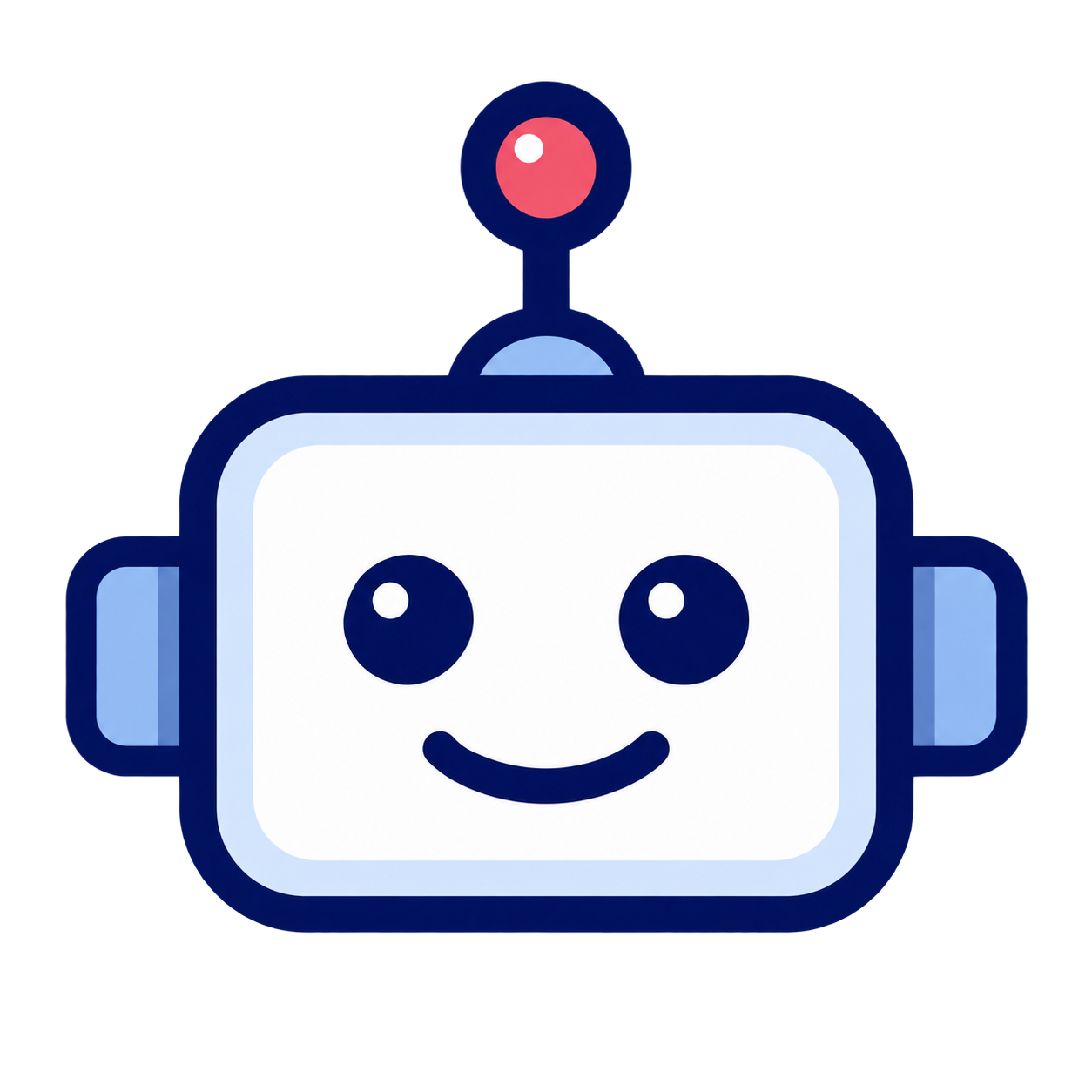}%
    }%
}
\begin{document}

\date{}

\title{\Large \bf Securing Web Agents with the Compile-Then-Execute Paradigm}

\author{
{\rm Annabella Chow}\\
UC Berkeley
\and
{\rm Jinhao Zhu}\\
UC Berkeley
\and
{\rm Sylvie Venuto}\\
UC Berkeley
\and
{\rm Abdulaziz Alharbi}\\
KACST
\and
{\rm Basel Alomair}\\
KACST
\and
{\rm Raluca Popa}\\
UC Berkeley
\and
{\rm David Wagner}\\
UC Berkeley
\and
{\rm Julien Piet}\\
UC Berkeley
} %

\maketitle

\input{sections/abstract}

\input{sections/introduction}

\input{sections/background}
\input{sections/problem_setup}
\input{sections/system_design}

\input{sections/evaluation}
\input{sections/discussion}

\input{sections/related_works}
\input{sections/conclusion}

\section*{Acknowledgments}
This research was supported by the KACST-UCB Joint Center on Cybersecurity and gifts from the Noyce Foundation, Google, Accenture, Amazon, AMD, Anyscale, Broadcom, Cisco, IBM, Intel, Intesa Sanpaolo, Lambda, Lightspeed, Mibura, NVIDIA, Samsung SDS, and SAP. We would like to thank Yiwei Hou, Muxi Lyu, Xiao Huang, Michael Lavery, Andy Joe for brainstorming ideas and students from Sky Security and David Wagner's Lab for their helpful feedback.

\bibliographystyle{plainurl}
\bibliography{references}
\clearpage
\raggedbottom
\input{sections/appendix}

\end{document}

%% file: sections/abstract.tex
\begin{abstract}

Web agents commonly follow the ReAct paradigm, selecting each action from runtime observations that inherently mix trusted and untrusted content.
This design makes their control flow vulnerable to prompt injections embedded in webpages.
We introduce \sys \footnote{Baya is named after the Baya Weaver, a Southeast/South Asian native bird known for building intricate hanging nests that protect eggs and chicks, a parallel to our compile-then-execute design, which constructs a secure plan before execution.}, a compile-then-execute web agent designed to prevent prompt-injection-driven control-flow hijacking by design.
Before accessing the live website, \sys{} compiles the user's request into an executable Python program that calls a Trusted Website API.
The planning model does not observe live website content; only the compiled program processes it.
Untrusted data may determine program values or select among branches defined during compilation, but it cannot redefine the user task or cause the model to synthesize new website operations at runtime.
The Trusted Website API gives the compiler a unified view of the website's action space, enabling ahead-of-time planning without the stepwise action discovery required by reactive agents.
Our evaluation across 581 valid WebArena Verified tasks shows that \sys{} achieves 77.8\% aggregate utility, outperforming browser-only ReAct (55.1\%) and API-enabled ReAct (66.7\%).
On WASP, a security benchmark for web agents, \sys achieves 0\% attack success rate, compared with 16.2\% for API-enabled ReAct and 25.5\% for browser-only ReAct.
\end{abstract}

%% file: sections/introduction.tex
\section{Introduction}
Web agents have emerged as a promising interface for automating complex web tasks, such as retrieving information, shopping online, or planning travel.
Modern web agents typically operate in a reactive loop: they observe the current webpage, reason about the next action, execute it through a browser interface, and repeat until the task is complete \cite{yao2023react}.
This ability to adapt to information revealed during execution enables agents to operate on dynamic websites and complete complex, multi-step tasks.

The same adaptability creates a fundamental security risk.
Unlike traditional software interfaces, webpages inherently contain a mixture of trusted and untrusted information, including content from the website itself, users, and third parties.
When a reactive agent incorporates this content into its reasoning, attacker-controlled content can influence not only the data it processes but also the actions it selects.
This opens the door to indirect prompt-injection attacks~\cite{debenedetti2024agentdojo, zhan2024injecagent}, in which malicious web content can cause the agent to deviate from the user's intended task and perform attacker-chosen actions~\cite{wang2025webinject}.
In this paper, we focus on preserving control-flow integrity: preventing untrusted web content from introducing attacker-chosen operations beyond the user's intended workflow.

Separating planning from execution and committing to a pre-compiled plan provides a structural defense against control-flow hijacking.
Instead of allowing the agent to continuously decide its next action based on runtime webpage content, the agent first compiles a task-specific plan and then carries out that plan against the web.
In this architecture, information retrieved from webpages is treated as runtime data consumed by a compiled plan, rather than as instructions that can arbitrarily redefine the agent's control flow.
Prior work, such as CaMeL and FIDES, has demonstrated that compile-then-execute is an effective foundation for secure tool-calling agents~\cite{debenedetti2025camel,costa2025fides,an2025ipiguard,kolluri2026optimizing}.
A prior analysis of WebArena found that all benchmark tasks are compatible with this paradigm, assuming complete trusted APIs~\cite{piet2026pte}.
However, applying compile-then-execute to web agents is more challenging than to conventional tool-calling agents.

First, compile-then-execute requires knowledge, before execution, of which operations are available and what they mean. This is typically available to tool-calling agents through predefined and documented tool interfaces. In contrast, web agents must interact with heterogeneous websites whose operations and semantics may be undocumented, incomplete, or difficult to determine in advance.
Most websites do not expose complete, documented, and semantically meaningful APIs for all actions available through their user interfaces.
As a result, there is a missing abstraction between today's web infrastructure and the stable semantic interface required by a compile-then-execute agent.

Second, even when trusted website APIs are available, there typically is little or no documentation of their semantics, dependencies, and site-specific rules. As a result, the agent must discover these rules, catch failures, diagnose their cause, and adapt the plan securely.
Past work typically focuses on ``one-shotting'' the plan, but we've found that web agents must revise and repair failing plans, which is challenging to do securely.

In this paper, we introduce \sys{}, a web agent based on the compile-then-execute paradigm that addresses these challenges for tasks whose workflows can be determined before execution.
During compilation, the agent emits an executable program, fixing its control flow and possible API interactions before live execution begins. Runtime data may fill program variables or select among branches defined during compilation, but it cannot introduce new task instructions.

\sys{} consists of three main components. First, we take a first step toward understanding what API abstractions are sufficient for web agents by manually constructing
a Trusted Website API that presents native REST endpoints and wrappers for UI-only functionality through a unified, documented interface augmented with site-specific execution constraints and prerequisites.
This interface is prepared for each website before user tasks are processed; automating its construction remains future work.
Second, a sandboxed coding agent compiles the user's task into a Python program and validates it using mocked tests without accessing the live website.
A catalog of available APIs, search tools over this catalog, and site-specific guidance help Baya find the relevant operations and write a program that uses them correctly.
Third, an isolated execution engine runs the program against the live website.
If execution fails, it exposes restricted diagnostic feedback to help the compiler repair the program.
Feedback is restricted to prevent it becoming a channel for attack.

By separating task compilation from runtime web interaction, \sys{} limits the ability of untrusted webpage content to alter the agent's intended control flow while retaining the ability to consume web data during execution. Outcome manipulation and general data-flow security are outside our scope.
Beyond security, \sys{} improves utility by providing a structured view of the website's available operations, reducing uncertainty about how to accomplish the user's task.

We evaluate \sys{} along three dimensions: utility, security, and efficiency. 
We measure task completion using WebArena Verified~\cite{hattami2025webarena},
a manually audited revision of WebArena~\cite{zhou2024webarena}.
We compare \sys to two simple ReAct agents (\reactApiWeb and \reactWeb) and to Untrusted Content Masking (UCM), a secure ReAct-style web agent that masks and quarantines untrusted content~\cite{nikolic2026ucm}, and evaluate latency and execution cost.
To assess security, we evaluate \sys{} on WASP~\cite{evtimov2025wasp}, measuring its robustness against prompt-injection attacks embedded in web content. 
Across four environments, \sys{} has the highest aggregate utility with all three models\introrev{, outperforming \reactApiWeb{} and \reactWeb{} by 11.1 and 22.7 percentage points, respectively. On GitLab, \sys{} also outperforms UCM by 17.6 percentage points.}
Moreover, \sys{} achieves 0\% observed intermediate and end-to-end ASR on WASP, compared to the ReAct baselines' aggregate ASRs of 16.2\% and 25.5\%, respectively.

\input{tables-figures/summary_table}

%% file: tables-figures/summary_table.tex
\begin{table}[t]
\centering
\small
\setlength{\tabcolsep}{6pt}
\renewcommand{\arraystretch}{1.1}
\captionsetup{width=0.92\columnwidth}

\caption{
Overview of utility, security, and efficiency.
Utility is averaged over four environments, security over two.
}

\begin{tabular}{@{}l
    *{3}{>{\centering\arraybackslash}p{1.8cm}}@{}}
\toprule
\textbf{System} &
\textbf{Utility\,$\uparrow$} &
\textbf{ASR\,$\downarrow$} &
\textbf{Latency (s)\,$\downarrow$} \\
\midrule
\sys         & \textbf{77.8\%} & \textbf{0.0\%} & 327.15 \\
\reactApiWeb & 66.7\% & 16.2\% & 86.60 \\
\reactWeb    & 55.1\% & 25.5\% & \textbf{59.91} \\
UCM\footnote{UCM is evaluated on GitLab environment only.}          & 75.3\% & \textbf{0.0\%} & 469.04 \\
\bottomrule
\end{tabular}

\label{tab:overview}
\end{table}

%% file: sections/background.tex
\section{Background}

Web agents use language models to translate a user's intent into actions that inspect or modify website state.
A web-agent design has two conceptually distinct dimensions: the interface through which the agent interacts with a website and the architecture through which it selects actions.

At the interaction layer, vision-based agents operate on screenshots and issue keyboard or pointer actions~\cite{he2024webvoyager,zheng2024seeact}.
DOM-based agents inspect structured page representations, such as the DOM or accessibility tree, and invoke browser actions on referenced elements~\cite{zhou2024webarena,chezelles2025browsergym}.
API-based agents invoke structured website operations without directly manipulating the browser interface~\cite{piet2026pte,song2023restgpt}, while hybrid agents combine API and browser access~\cite{song2025beyond, zhang2025apivsgui}.
These interfaces determine how an action is expressed, but they do not determine when or how the agent selects that action.

At the control layer, a common architecture is ReAct, which interleaves model reasoning with environment interaction~\cite{yao2023react}.
At each step, the model receives the user goal and the latest environment observation, selects one action, executes the action, and observes the result before selecting its next action.
Consequently, the agent's future actions and control flow remain unspecified until runtime observations become available.
This property applies regardless of whether the observation originates from a screenshot, a structured page representation, or an API response.

\subsection{Security Challenges}
\label{ssec:security-challenges}

The web is a mixed information environment.
A single webpage often combines trusted, site-provided interface elements with untrusted content from users or third parties, such as product reviews, forum posts, advertisements, and embedded widgets. 
The trust boundary is often ambiguous, making it difficult for agents to tell apart trusted from untrusted information. 
Even if they could, ignoring untrusted data would sacrifice utility; for example, an agent tasked with summarizing product reviews inherently needs to read user-supplied text to succeed.

This mixed environment creates a security flaw for ReAct-based agents. 
Because ReAct determines its control flow dynamically, untrusted web content is presented to the language model exactly when it is deciding on its next action. 
By collapsing observation and action selection into a single inference step, the agent's decision process can be hijacked by untrusted content through indirect prompt injection attacks.

We distinguish between two primary goals an attacker might pursue via prompt injection: control-flow hijacking and outcome manipulation. \sys{} protects against control-flow hijacking, outcome manipulation is out of scope. 

\noindent\textbf{Control-flow hijacking} is the most severe threat, analogous to arbitrary code execution. 
Because a ReAct agent continuously generates arbitrary tool calls on the fly, an attacker can use prompt injection to make the model deviate from the user's intent and synthesize entirely new, unauthorized actions at runtime. 
For example, malicious text in a forum post could instruct an agent to navigate to a settings page and exfiltrate the user's private data. 
This is a ``confused deputy'' vulnerability: because web agents often act with the user's logged-in authority, injected instructions can trick the privileged agent into using its authority on behalf of the attacker. The more tools and permissions the agent has access to, the more reach the attacker has. 

\noindent\textbf{Outcome manipulation}, on the other hand, is closer to a data-only attack. Here, the attacker influences the outcome of the user's task without changing the task itself or forcing the agent to invoke unintended tools. For instance, an adversarial seller might use injected text to trick an agent into purchasing a counterfeit product instead of the legitimate item the user requested. 

The distinction is whether untrusted content can expand or redefine the agent's workflow.
Outcome manipulation changes runtime values or selects among branches already permitted by the workflow, whereas control-flow hijacking introduces attacker-directed operations or branches. 

\subsection{The Architectural Gap}
Several existing defenses attempt to secure the ReAct loop by hardening the underlying language models to ignore malicious instructions, by wrapping the agent in runtime monitors to block unauthorized actions, or by having the agent write a high-level plan before proceeding with a ReAct loop. 
These methods fail to address the core architectural flaw: interleaving planning with untrusted observations~\cite{meng2025cellmate, zhang2025browsesafe, wallace2024instruction, wang2025agentarmor}.

A recent line of work in structural defenses, such as CaMeL~\cite{debenedetti2025camel}, attempts to remove this vulnerability by strictly separating planning from execution. In this paradigm, a language model converts the trusted user query into a program before any interaction with the environment occurs, and this code is then executed by an interpreter. Because the control flow is fixed as a static program before runtime data is observed, untrusted content is confined to variables. It cannot modify the program's execution path or synthesize new actions.

However, these plan-then-execute approaches have so far only been applied to simple agentic environments with a small, well-defined action space~\cite{debenedetti2024agentdojo,debenedetti2025camel,
costa2025fides,an2025ipiguard}. In these settings, the agent has full visibility into the available tools, and the side effects of each action are known prior to execution. 

Existing web interfaces do not naturally provide this abstraction.
Low-level browser actions, such as clicking or typing, derive their meaning from the current rendered page, and many actions become visible only after earlier interactions.
Native backend APIs provide more stable operations, but they are frequently incomplete, undocumented, or inconsistent with functionality exposed through the user interface.
Consequently, neither browser actions nor raw backend APIs directly provide the complete, semantically specified action space required for compile-then-execute.

%% file: sections/problem_setup.tex
\section{Problem Setup}

\begin{figure*}[t]
    \centering
    \includegraphics[width=\linewidth]{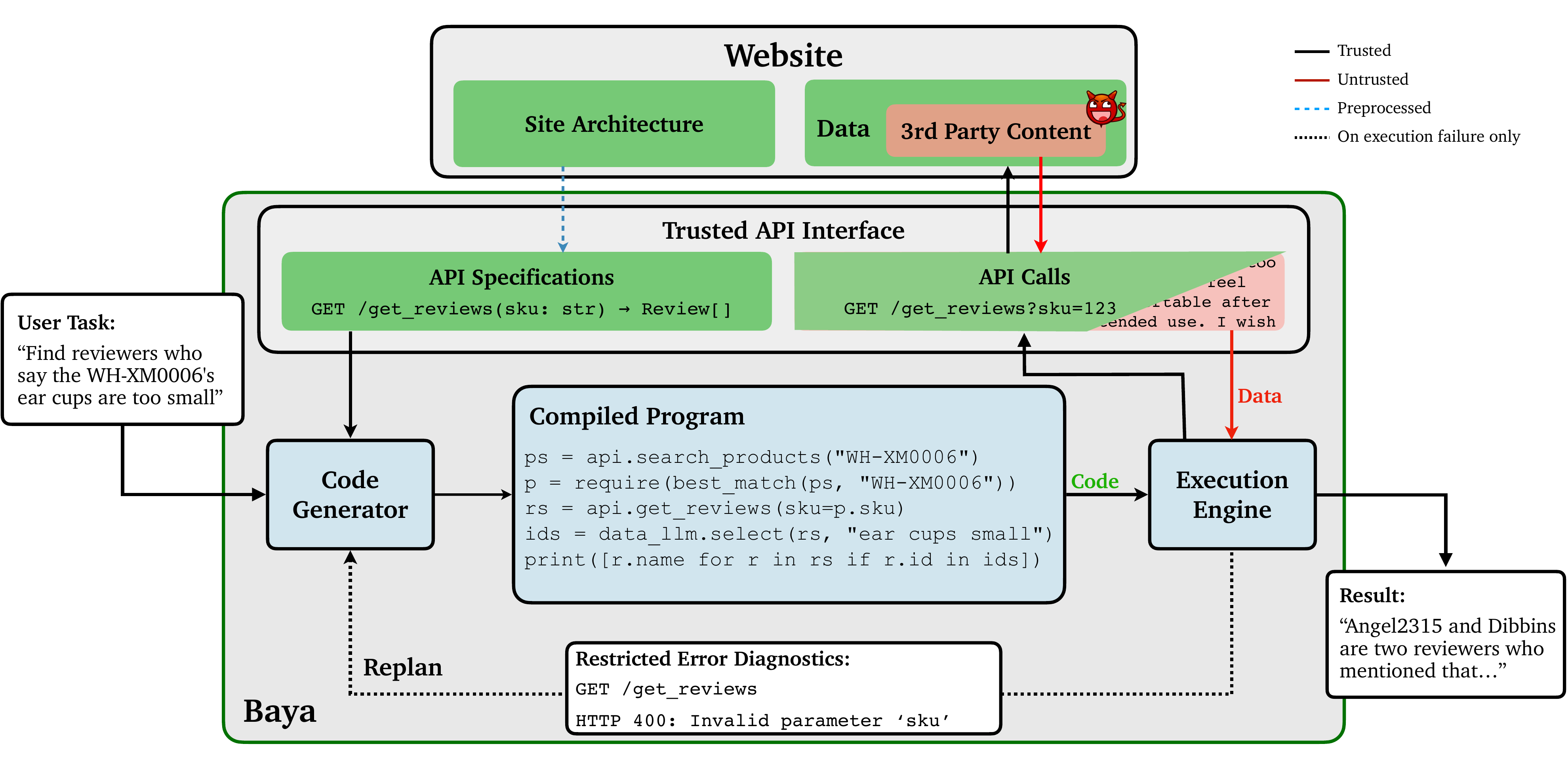}
    \caption{\textbf{Architecture of \sys.}
    \sys generates an executable program using trusted API
    specifications without accessing website data.
    During execution, untrusted data is consumed by the
    program but cannot alter its precompiled control-flow
    structure. Replanning occurs only upon execution failure,
    with only restricted error diagnostics and the existing
    code passed back to the Code Generator, without exposing
    any website data.}
    \label{fig:arch}
\end{figure*}
\subsection{Goals}

Our design pursues two goals: preserving the control-flow integrity of the user's task and maintaining utility on user tasks.

\noindent\textbf{Security.}
Our primary security goal is to completely prevent control-flow hijacking. 
An attacker must not be able to override the user's original intent or synthesize new, unauthorized actions at runtime.
The operations available to the agent and their control dependencies must be determined exclusively from the trusted user request and trusted website interface before live execution begins.
Untrusted runtime data may instantiate program variables and select among branches defined during compilation, but it must not introduce new operations, execution branches, or tasks.

Our security objective is limited to preventing control-flow hijacking.
Outcome manipulation, misinformation, false information, denial-of-service attacks, and compromises of trusted components are outside our scope.

\noindent\textbf{Utility.}
We also seek to build a useful web agent that can execute user tasks. A fundamental limitation of compile-then-execute is that it cannot support tasks where the workflow is not determined by the user task (e.g., ``open my email inbox, find the email from my boss, and do whatever it says to do'').
To evaluate how many user tasks are compatible with compile-then-execute, we classify web tasks into one of three categories:
\begin{itemize}[leftmargin=*,noitemsep,topsep=2pt]
    \item \textbf{Safe.} The complete task can be expressed as a deterministic program that processes runtime data without invoking a language model.
    For example, adding the highest-rated noise-canceling headphones under \$200 to a cart requires retrieval, sorting, and a predetermined cart operation.

    \item \textbf{Safe with Influence.} The website operations and control-flow structure can be fixed before execution, but the task may require a runtime model to interpret unstructured data.
    For example, summarizing product reviews requires semantic processing, but the runtime model need not invoke tools or modify the compiled workflow.

    \item \textbf{Replan-Needed.} The operations required to complete the task must be derived from untrusted runtime content.
    For example, a request to read incoming emails and complete every action item requires the agent to treat untrusted email content as a source of new tasks.
\end{itemize}
Compile-then-execute can securely support both Safe and Safe with Influence, but cannot support Replan-Needed tasks, which are incompatible with the restrictions needed for security.

A prior analysis~\cite{piet2026pte} of WebArena~\cite{zhou2024webarena} classifies 81\% of WebArena tasks as Safe, 19\% as Safe with Influence, and none as Replan-Needed.
This suggests that compile-then-execute can support many web tasks.

\subsection{Threat Model}

We focus on indirect prompt-injection attacks against web agents.
The user assigns the agent a benign web task, and the attacker embeds malicious instructions in content that the agent may process while completing that task.

We consider third-party attackers, such as commenters, reviewers, forum users, third-party sellers, and advertisers.
We assume that website operators and model providers are trusted and do not participate in the attack.
Accordingly, the website backend and the code implementing its APIs are part of our trusted computing base.
We assume that these components correctly implement the website's intended functionality.

In contrast, content originating from website users or other third parties is untrusted.
This content remains untrusted whether the agent retrieves it through a rendered webpage, a native website API, or a browser-backed API wrapper.
We treat any such content encountered during live execution as potentially attacker-controlled.

%% file: sections/system_design.tex
\section{System Design}
\sys\footnote{https://github.com/apechow/Baya} processes web tasks in two isolated phases (\cref{fig:arch}). During compilation, a sandboxed coding harness generates and validates a Python program using the Trusted Website API, without accessing live website content. During execution, a separate engine runs the program against the live website.

\subsection{Constructing the Trusted API}
\label{sec:trusted-api}
Central to this architecture is the Trusted Website API, which acts as the core abstraction enabling ahead-of-time planning. We require the Trusted Website API to be functionally complete over the target task space, strictly typed to support static code generation, and richly documented so that \sys{} can determine the semantic meaning of each operation prior to execution.

\sys{} does not assume that such an interface is already provided by the website. Thus, before \sys{} processes user tasks, we construct a Trusted Website API for each target website that exposes the operations the agent may need. This construction is part of the trusted, offline setup of the system: once the interface is fixed, \sys{} plans only over these predefined operations and cannot introduce new website capabilities at runtime. Consequently, replanning can correct how an existing operation is used, but cannot recover an operation that is absent from the Trusted Website API.

To construct this interface in WebArena, we begin with the website's native backend REST APIs, which provide structured access to much of the site's functionality. \revision{We manually inspect each website's backend and available API documentation to identify its supported operations,} extract their schemas, normalize inconsistencies in their request and response specifications, and augment them with missing semantic constraints needed for correct use. Native APIs, however, do not cover all functionality exposed through the website UI. \revision{Through manual inspection of the website's functionality, we identify gaps in the available backend operations and fill them with trusted browser-automation operations implemented using Playwright. These additional operations are exposed through the same typed interface, providing sufficient coverage of the website's functionality for agents to accomplish user tasks without direct browser interaction.} The resulting Trusted Website API is thus a curated combination of native backend operations and additional trusted operations.

Constructing this interface is nontrivial because real websites expose heterogeneous schemas, omit execution-critical constraints, and often lack backend operations corresponding to UI functionality. Below, we use the WebArena websites as a case study to characterize these challenges and then describe the contract we impose on every operation in the Trusted Website API.

\subsubsection{Challenges in Real-World Website APIs}

\heading{Challenge 1: Heterogeneous and Underspecified Schema Dialects.}
Even among REST APIs, request formats and parameter conventions differ substantially across websites. For example, searching and filtering on the Shopping site requires a specific nested bracket parameter format:
\nolinkurl{searchCriteria[filterGroups][0][filters][0][conditionType]}.
The composition semantics are not self-explanatory: filters with the same \texttt{filterGroups} index are combined using \texttt{OR}, whereas filters with different indices are combined using \texttt{AND}. Furthermore, although Shopping and Shopping Admin share the same Magento backend, they assign different semantics to identical features: item variants are represented as option IDs in Shopping, but as distinct child products with individual SKUs in Shopping Admin. Additionally, websites use conflicting conventions for request bodies: GitLab describes parameters using \texttt{formData} or a generic \texttt{body} field, whereas Shopping uses auto-generated parameter names that do not match actual JSON payload keys, hindering valid request serialization. Representative API schema excerpts are provided in
\cref{fig:api-schema-examples}.

\heading{Challenge 2: Incomplete Descriptions and Undocumented Constraints.}
Schemas frequently omit constraints required for successful execution. For example, GitLab exposes extensive REST APIs but omits response descriptions. Shopping's add-to-cart documentation marks product options as optional. In reality, configurable products (such as clothing or shoes) fail unless specific size and color options are supplied. Because option types and valid values vary per product, the agent must first inspect the product's available options before constructing the add-to-cart mutation. Such dependencies and identifier formats (e.g., whether an endpoint expects a numeric user ID or a username string) are absent from raw endpoint schemas.

\heading{Challenge 3: Incomplete Task Coverage and UI Semantic Disconnect.}
Backend REST APIs frequently fail to cover the full range of operations supported by website user interfaces. For example, WebArena's Postmill forum supports only 16 REST APIs, which are enough for only 33\% of its tasks~\cite{piet2026pte}; operations such as voting or replying to a forum have no REST counterpart. In GitLab, administrative actions such as generating an RSS token are UI-only. Similarly, Shopping Admin reports on popular search terms and best-selling items are accessible exclusively through the admin dashboard. Furthermore, REST endpoints for search often support only exact-match searches, whereas the UI supports fuzzy search. For instance, Shopping's product-search API requires the exact product title, failing on descriptive queries such as ``reorder the makeup removal kit I bought in summer 2022'', which the storefront search UI resolves natively.

These coverage gaps cannot be resolved by schema normalization alone because the required backend operation simply does not exist. During construction of the Trusted Website API, we therefore identify task-relevant UI functionality not covered by native APIs and implement corresponding trusted Playwright operations. These operations expose the same high-level semantics as native API operations while hiding low-level browser interaction from \sys{}.

\subsubsection{API Contracts}
To bridge these gaps, \sys structures its action space as a \emph{Trusted Website API}. Formally, we define a Trusted Website API as a collection of callable operations $\mathcal{A} = \{f_1, f_2, \dots, f_n\}$, where each operation $f \in \mathcal{A}$ is governed by a machine-readable contract $\mathcal{C}(f) = (\Sigma, \mathcal{P}, \mathcal{Q}, \mathcal{E})$ structured as:

\begin{enumerate}[leftmargin=*,itemsep=1pt,topsep=2pt]
    \item \textbf{Type Signature ($\Sigma$):} A function signature $f: \mathcal{X}_1 \times \dots \times \mathcal{X}_k \to \mathcal{Y}$. All input parameters $\mathcal{X}_i$ and return structures $\mathcal{Y}$ are defined as strongly typed records (e.g., Python dataclasses) with explicit primitive types, structural constraints, and field-level docstrings.
    \item \textbf{Preconditions ($\mathcal{P}$):} Constraints on input states required prior to invocation. This includes authentication token availability, session requirements, or prerequisite queries (e.g., querying the available stock of an item before adding it to the cart), and API-specific parameter conventions.
    \item \textbf{Postconditions \& Side Effects ($\mathcal{Q}$):}
    What the function returns and how it changes the state of the website
    upon successful invocation. Contracts explicitly categorize operations
    into \emph{idempotent reads} (pure data retrieval) and
    \emph{stateful mutations} (e.g., order submission, entity creation).
    
    \item \textbf{Error Domain ($\mathcal{E}$):} An exhaustive declaration of structured error conditions (e.g., entity not found, invalid parameter, authentication expired) mapped to typed exceptions, ensuring that partial failures do not emit unstructured or untyped side effects.
\end{enumerate}

This contract directly overcomes the three practical challenges. By canonicalizing request parameters and response schemas into strongly typed signatures $\Sigma$, the contract eliminates the dialect heterogeneity of raw HTTP interfaces (Challenge~1). By encoding prerequisite lookups, identifier formats, and execution constraints within preconditions $\mathcal{P}$ and side-effect guarantees within postconditions $\mathcal{Q}$, the specification provides known execution requirements upfront (Challenge~2). Finally, the contract establishes a uniform interface whose underlying implementation is chosen based on backend availability: an operation is implemented via a backend REST API when available, or via Playwright browser automation when functionality is exposed exclusively through the user interface or requires fuzzy storefront search (Challenge~3). Because both implementations satisfy the exact same signature $\Sigma$ and return structure $\mathcal{Y}$, the planner targets a complete, consistent action space without exposing the planning process to DOM-level mechanics.

\subsubsection{Worked Example: Product Search Endpoint}
\input{tables-figures/trusted_api_example}

Figure~\ref{fig:trusted-api-example} illustrates this architecture with a concrete worked example for an e-commerce product search endpoint.

\heading{Contract Specification.}
As shown at the top of Figure~\ref{fig:trusted-api-example}, the operation \texttt{search\_products} defines a formal contract returning a list of \texttt{ProductListing} objects. The return structure $\mathcal{Y}$ specifies exact fields: \texttt{sku}, \texttt{name}, \texttt{price}, \texttt{in\_stock}, \texttt{rating}, and \texttt{url}. The preconditions $\mathcal{P}$ require non-empty queries and valid authentication. The postconditions $\mathcal{Q}$ certify that the operation is an idempotent read with zero server-side mutations \revision{and returns an empty list \texttt{[]} when no products match}, while the error domain $\mathcal{E}$ maps HTTP 401/403 to \texttt{AuthenticationError} and returns an empty list \texttt{[]} when no products match.

\heading{Implementation Alternatives.}
The lower half of Figure~\ref{fig:trusted-api-example} illustrates two alternative implementations of this contract depending on backend availability:
\begin{itemize}[leftmargin=*,itemsep=1pt,topsep=2pt]
    \item \textbf{REST Implementation (Left):} Serializes the query into Magento's complex nested bracket parameter structure (\texttt{\seqsplit{searchCriteria[filterGroups]...}}), issues a \texttt{\seqsplit{GET /rest/V1/products}} request with authorization headers, checks response status, and maps raw JSON dictionary keys into typed \texttt{ProductListing} dataclass instances.
    \item \textbf{Playwright Implementation (Right):} Launches a headless browser, navigates to the storefront search URL (\texttt{\seqsplit{/catalogsearch/result/?q=...}}), locates rendered product cards via DOM selectors (\texttt{\seqsplit{.product-item-info}}), parses price strings and stock availability indicators, and returns a list of \texttt{\seqsplit{ProductListing}} instances identical in structure and type to the REST binding.
\end{itemize}

During compilation, \sys writes and tests its program against the abstract \texttt{ProductListing} contract using mocked unit tests, targeting a uniform typed interface regardless of whether a given function executes via REST or Playwright.

\heading{API Construction Limitations.}
While we manually construct Trusted Website APIs for the fixed WebArena environments, extending \sys{} to arbitrary websites requires scalable methods for discovering and validating trusted operations.
Existing systems such as Libretto demonstrate how browser interactions can be converted into reusable workflows~\cite{libretto-browser-workflows}.
However, automating this process introduces a new security challenge: prompt injections encountered during website exploration could influence operations subsequently treated as trusted.
We discuss this further in~\cref{sec:missing_api_capabilities}.

\subsection{Compile-then-Execute}
To complete web tasks without exposing the planning process to untrusted live-site content, \sys separates program generation from live execution. It uses a two-phase Compile-then-Execute architecture. During compilation, a coding harness connected to an LLM translates the task into an executable program using the provided API information, site-specific rules, and output requirements. During execution, a separate execution engine runs the generated program against the live site. The two phases communicate through generated artifacts and restricted error diagnostics rather than through an interactive session.

\subsubsection{Compilation}
Because \sys cannot probe the live site, it must receive enough information to construct a working program in advance. At the beginning of each task, the agent receives the task intent, starting project path, account identity, authentication instructions, output requirements, and Trusted Website API information. These specifications include a complete endpoint catalog, tools for searching and inspecting API contracts, and site-specific execution requirements. For each operation, the contract specifies its signature ($\Sigma$), preconditions ($\mathcal{P}$), such as prerequisite lookups and identifier resolution, postconditions and side effects ($\mathcal{Q}$), and structured error conditions ($\mathcal{E}$).

\heading{API Discovery.} 
\sys's API discovery is efficient and scalable, avoiding the context-window costs of loading complete API schemas upfront. Loading an entire website's API schema into an LLM's context window is impractical. A single website may expose hundreds or thousands of operations, and loading schemas for additional websites
upfront causes context consumption to grow accordingly. For example, GitLab exposes 1,028 executable REST APIs, and its schema file size is approximately 1.8 MB.
Loading the complete schema would consume a substantial portion of the agent's context window before it begins reasoning about the task.

To address this challenge, \sys provides several levels of API discovery without requiring the agent to read the complete schema. \revision{Site-specific execution requirements} describe common pitfalls and preferred endpoint choices, while a compact endpoint catalog lists all available paths and methods. The agent can then use \sys's curated
tools to search for relevant endpoints and retrieve detailed schemas only for the selected operations. Anthropic's Tool Search Tool follows a similar high-level principle, avoiding loading all tool definitions upfront and instead discovering and expanding only those relevant to the current task \cite{wu2025advancedtooluse}.

\heading{Reliable program construction.}
Discovering the relevant API operations is only one part of compilation; the agent must also compose them into a program that executes reliably within its resource budget. 

\revision{Efficient retrieval presents a tradeoff between underconstrained searches that retrieve too many candidates and overly restrictive filters that discard valid results. For example, a generic product search on the Shopping website may fetch and inspect up to 100 pages, potentially requiring additional API calls for every product on each page. Such searches can quickly exhaust the execution-time budget. Server-side filters and deterministic local matching can reduce this cost, but they may be insufficient for tasks requiring semantic judgment. For example, a task asking for reviewers who mention that a product's ear cups are too small cannot be solved reliably by matching only the word ``small,'' since reviewers may describe the same issue using different words or phrases.

To balance these competing concerns, we prompt \sys to use server-side filters, bound pagination, and apply deterministic matching utilities whenever possible. For tasks requiring semantic judgment, the generated program may instead include a narrowly scoped runtime LLM call after deterministically narrowing the candidate set. This approach aims to reduce unnecessary API calls while preserving flexibility for semantic matching, although both underconstrained retrieval and overly restrictive filtering remain possible. We examine these failure modes in our evaluation.}

\input{tables-figures/pseudocode}
\heading{Generated artifacts.} 
\sys produces two artifacts in the compilation phase: a task-specific Python program and a companion unit-test file (e.g., \cref{fig:generated-program-pseudocode} and \cref{fig:generated-test-pseudocode}). The generated program is a restricted Python script: it uses the Python standard library for local computation, invokes only the trusted APIs for website access, and does not use arbitrary code execution or unrestricted network access. Retrieval, filtering, counting, arithmetic, sorting, and exact or fuzzy matching are implemented using deterministic code. When a task requires semantic interpretation of free-form text that cannot be expressed reliably using deterministic logic, the program may include a narrowly scoped, data-only LLM call. Such a call operates only on candidate data already retrieved and narrowed through trusted APIs; it cannot retrieve additional data or perform actions on the website. During compilation, \sys has no access to the live website and validates the program entirely against local mocks. The companion test file mocks all HTTP requests, authentication and credentials, and any LLM client used by the program, and tests both successful execution and expected error paths.~\cref{app:example-artifacts}~\cref{fig:toy-generated-workflow} illustrates a toy program and artifacts that \sys{} might generate.

\subsubsection{Design Guarantees and Restrictions}

The security of the design depends on preventing live-site data from influencing the program while it is being generated. Therefore, compilation and execution are separated in time, process, privilege, and information. Code is generated before live execution begins, and the coding harness and execution engine run as separate processes. The compilation process receives only the information and permissions needed to generate and test the program, whereas the execution process receives the runtime domains and credentials needed to interact with the live site.

To enforce this boundary, \sys can read only the provided task context, approved API schemas and supporting tools, and files in its assigned task directory. It can write only to that task directory. 
It cannot access any websites, the network, web search, credentials, evaluation tests, coding-harness memory, Git metadata, or unrelated files. Consequently, the agent must construct the program using the task and trusted API information rather than information retrieved from the web or live site.

These restrictions are enforced through multiple layers, including coding-harness permissions, tool restrictions, an OS-level filesystem sandbox, a network sandbox, and explicit prompt instructions. The concrete harness settings, filesystem allowlists, permitted commands, and network-sandbox configuration are provided in~\cref{app:example-artifacts}~\cref{fig:coding-harness-configuration}.

\subsubsection{Validation}
\label{sec:validation}

The goal of validation is not to guarantee that the program will avoid every runtime failure, but to verify that it detects and reports anticipated classes of failures correctly. In-scope conditions include non-\texttt{2xx} API responses; empty result sets; missing, null, or incorrectly typed parameters or response fields; operations over empty inputs; and unparseable dates or numbers. The tests also verify that identifiers are resolved from mocked runtime responses rather than hardcoded and, when the program uses an LLM for filtering, that the call uses the configured model and correctly handles a ``none of these'' result.

For example, suppose a task asks \sys{} to add a clothing item to the user's wishlist. The API schema marks the quantity parameter as optional, so the compiled program may omit it, even though the live \texttt{/wishlist} endpoint actually requires it and returns an HTTP 500 response when it is absent. Validation cannot infer this undocumented requirement or prevent the runtime failure. It can, however, verify that if an API call returns a non-\texttt{2xx} response, the program detects the error, reports relevant diagnostic information, and exits with an error code rather than producing a final answer. During live execution, this error is then passed back to \sys{}, which can regenerate a new plan that accounts for the previously undocumented requirement.

Thus, unit testing verifies the program's behavior under anticipated conditions without performing the task or interacting with the live site. Failures that depend on undocumented or otherwise unobserved live-site behavior are discovered during execution and handled through replanning.

\subsubsection{Execution}
After validation, a separate execution engine receives the generated program and runs it against the live website in a Python subprocess.
The execution engine itself simply executes Python code, which retrieves data or modifies website state through trusted API
operations. 
The code itself may contain LLM calls which are only used for data processing, producing no side effects and returning typed variables. During execution, the program records its progress in a Python output log. If execution succeeds, the program returns its final answer. Otherwise, \sys may replan (generate a new plan, based on the received error code) and retry the program up to three times.

\subsubsection{Error Handling and Replanning}
Local validation cannot anticipate every failure that may arise when a generated program interacts with the live website, as discussed in~\cref{sec:validation}. 
When execution fails, \sys{} may replan using the previously generated program together with the restricted error diagnostic. For example, a failed product query may return:
\begin{quote}
\begin{verbatim}
ERROR search_products returned HTTP 400
ERROR Response body:
{"message": "Invalid filter field: category"}
ERROR Product search failed
\end{verbatim}
\end{quote}
This feedback allows the compiler to revise the program, for example by replacing an unsupported \texttt{category} filter with a supported query parameter.

This error channel does not expose arbitrary website content to the planning model. The program itself was generated before execution and therefore contains no content obtained from the live website. In particular, non-\texttt{2xx} error responses may contain backend-generated diagnostics such as invalid-parameter names, missing required fields, authentication failures, or malformed requests, but they must not contain or reflect stored resources, successful response values, request values derived from website content, or other user-controlled website data. Thus, an endpoint may report that \texttt{quantity} is required, but it cannot include a product description, review, username, post body, or other content obtained from the website.

This restriction is necessary because values manipulated by the generated program may themselves originate from prior API responses. \sys{} therefore does not rely on whether a value appears in a request or an exception to determine whether it is safe to expose during replanning. Instead, the replanning interface admits only diagnostics explicitly generated by the trusted API implementation; runtime values and website-derived data are excluded. An API operation that reflects attacker-controlled or website-derived content into its error message would violate the Trusted Website API contract and is outside our threat model.

Only an observable failure triggers another compilation attempt. During replanning, the agent receives the previous program and the restricted error diagnostics, then revises the implementation accordingly. Replanning stops once the program successfully returns an answer or after at most three replans. \revision{The program distinguishes legitimate empty results from execution failures by reporting the number of results found without exposing their contents, while API failures produce error diagnostics without exposing website data.} Legitimate empty results may be returned as ``not found,'' whereas authentication failures, rate limits, server errors, timeouts, parsing failures, and other execution errors remain failures.

%% file: tables-figures/trusted_api_example.tex
\begin{figure*}[t]
\centering
\small
\raggedright
\begin{lrbox}{\trustedapibox}
\begin{tcolorbox}[
    enhanced,
    title={RFC-Style Contract Specification:
    \texttt{search\_products}},
    colback=white,
    colframe=black,
    colbacktitle=black,
    coltitle=white,
    fonttitle=\bfseries\small,
    boxrule=0.4pt,
    arc=1.5mm,
    left=0.7mm,
    right=0.7mm,
    top=1mm,
    bottom=1mm,
    toptitle=0.7mm,
    bottomtitle=0.7mm,
    before skip=0pt,
    after skip=0pt
]
\begin{itemize}[
  leftmargin=*,
  itemsep=0pt,
  topsep=1pt,
  parsep=0pt,
  partopsep=0pt
]
  \item \textbf{Signature ($\Sigma$):}
    \texttt{search\_products(query: str) -> List[ProductListing]}

  \item \textbf{Data Structure ($\mathcal{Y}$):}\\[-0.3ex]
    \texttt{ProductListing \{sku: str, name: str, price: float, in\_stock: bool, rating: Optional[float], url: str\}}%

  \item \textbf{Preconditions ($\mathcal{P}$):}
    \texttt{len(query) > 0}; valid auth token for protected catalogs.

  \item \textbf{Postconditions ($\mathcal{Q}$):}
    Returns matching listings ordered by relevance. Idempotent read; no state mutation.

  \item \textbf{Error Domain ($\mathcal{E}$):}
    Raises \texttt{AuthenticationError} on 401/403; returns \texttt{[]} on zero matches.
\end{itemize}

\vspace{1mm}
\hrule height 0.4pt
\vspace{1.5mm}

\noindent

\begin{minipage}[t]{0.49\linewidth}
\vspace{0pt}
\begin{toycode}{1. Backend REST Implementation}
search_products(min_price, max_price):
    params <- {
    filterGroups[0].filters[0] = 
        price >= min_price
    filterGroups[0].filters[1] = 
        price <= max_price
    }

    response <- GET /products with params
    return response.items as ProductListing

\end{toycode}

\end{minipage}
\hfill
\begin{minipage}[t]{0.49\linewidth}
\vspace{0pt}
\begin{toycode}{2. Headless Playwright Implementation}
search_products(query):
    navigate to search page with query
    results <- displayed products

    for each result:
        extract sku, name, price

    return results as ProductListing
\end{toycode}
\end{minipage}
\end{tcolorbox}
\end{lrbox}

\resizebox{0.96\textwidth}{!}{%
    \usebox{\trustedapibox}%
}
\vspace{1.5mm}
\caption{Example Trusted Website API contract and its REST and Playwright implementations for an e-commerce product search endpoint.}
\label{fig:trusted-api-example}
\end{figure*}

%% file: tables-figures/pseudocode.tex
\begin{figure}[t]
\centering
\small
\caption{Pseudocode of a program generated by \sys{} to update a user's biography on the discussion forum.}
\begin{toycode}{Generated Program (Pseudocode)}
USERNAME <- "MarvelsGrantMan136"
NEW_BIO <- "Freelance Web Developer"

update_biography():
    response <- POST /update_biography {
    username: USERNAME,
    biography: NEW_BIO
    }

    if request failed or
       response.success is false:
        report error and exit

report success
\end{toycode}

\vspace{1mm}

\label{fig:generated-program-pseudocode}
\end{figure}

\begin{figure}[t]
\centering
\small
\caption{Pseudocode of generated unit tests that validate successful execution,
anticipated failures, and final-answer formatting without contacting the live
service.}
\begin{toycode}{Generated Tests (Pseudocode)}
test_success():
    mock success
    run update_biography()
    assert correct request and success

test_failures():
    for failure in [HTTP error, API failure,
    invalid JSON, exception]:
        mock failure
    assert update_biography() exits

test_final_answer():
    mock success
    output <- main()
    assert output contains NEW_BIO
\end{toycode}

\vspace{1mm}

\label{fig:generated-test-pseudocode}
\end{figure}

%% file: sections/evaluation.tex
\section{Evaluation}
\label{sec:evaluation}
We structure our evaluation around three dimensions: utility, security, and efficiency. First, we measure how reliably \sys{} completes benign web tasks relative to reactive baselines. We then evaluate whether \sys{} prevents control-flow hijacking while preserving utility on the intended task. Finally, we measure the computational and latency overhead introduced by \sys{} relative to the evaluated baselines.

We evaluate \sys{} and the ReAct baselines using three model families: \texttt{\seqsplit{claude-sonnet-4.6}}~\cite{anthropic-claude-sonnet-4-6}, \texttt{\seqsplit{gemini-3.5-flash}}~\cite{google-gemini-3-5-flash}, and \texttt{\seqsplit{gpt-5.6-luna}}~\cite{openai-gpt-5-6-luna}. For security evaluation, we replace \texttt{\seqsplit{gpt-5.6-luna}} with \texttt{\seqsplit{deepseek-v4-flash-0731}}~\cite{deepseek2026deepseekv4}. See~\cref{app:experimental_setup} for more details.

\subsection{Benchmarks}

\noindent\textbf{Utility}: WebArena ~\cite{zhou2024webarena} evaluates agents on realistic tasks across multiple self-hosted websites.
Each task specifies a natural-language intent, one or more starting URLs, and an evaluator that determines whether the task was completed successfully.
We evaluate the four environments supported by \sys{}: GitLab, Shopping, Shopping Admin, and Postmill.
We use a subset of the revised task intents and corrected ground truth from WebArena Verified~\cite{hattami2025webarena}, excluding tasks that have vague intents or cannot be evaluated due to the inherent nature of the task (i.e. drafting a message without submitting the form). 
We adapt the WebArena evaluation suite to score API-based agents by task outcome rather than browser-specific interaction traces. See~\cref{app:evaluation-method} for more details.

\noindent\textbf{Security: }
We evaluate resistance to prompt-injection-driven control-flow hijacking using WASP~\cite{evtimov2025wasp}.
We evaluate 12 GitLab and 9 Postmill attack templates, each paired with two benign user goals and two injection formats, producing 48 GitLab and 36 Postmill attack instances.
We report intermediate attack success rate (ASR), end-to-end ASR, and benign-task utility.

\subsection{Baselines}

We compare \sys{} against two reactive agent configurations and one security-oriented web agent.

\noindent\textbf{ReAct baselines.}
We implement a ReAct agent that alternates between model inference and environment interaction.
At each step, the model observes the current environment state, reasons about the task, and emits one action.
The resulting observation is returned to the model before it selects the next action.
This process continues until the agent completes the task, enters a repeated-action loop, or reaches the 30-step limit.
We evaluate two tool configurations.
\reactWeb interacts with websites exclusively through browser actions.
\reactApiWeb can use both browser actions and API calls, and receives the same API schemas and site-specific guidance as \sys{}.
Comparing \sys{} with \reactApiWeb isolates the effect of compile-then-execute from the effect of providing structured website interfaces.

\noindent\textbf{Untrusted Content Masking.}
We compare \sys{} with Untrusted Content Masking (UCM)~\cite{nikolic2026ucm}, a security-oriented browser agent that masks untrusted page regions from the primary agent.
When the agent must inspect a masked region, UCM routes the content through a quarantined model whose output is restricted by type.
Because UCM requires site-specific masking rules, we evaluate it only on GitLab, for which the published implementation provides the necessary rules.
We allow UCM up to 120 steps to align with their original evaluation, and we use Claude Sonnet-4.5 for the quarantined model.

\noindent\textbf{Other baselines.}
We also considered ceLLMate~\cite{meng2025cellmate} and Single-Shot Planning for Computer Use Agents (CUAs)~\cite{foerster2026camelscua}, two defenses against control-flow hijacking attacks.
However, neither system can be evaluated directly on WebArena Verified using its published experimental setup.
Like \sys{}, ceLLMate relies on site-specific artifacts prepared before deployment; specifically, it uses a browser-level sandbox to enforce security policies assumed to be provided by each website.
Because the published implementation does not include policies for WebArena Verified, evaluating ceLLMate would require us to construct additional benchmark-specific policies, precluding a direct comparison with the published system.
CUA's implementation and evaluation target OSWorld and do not provide an integration with WebArena Verified.
We therefore could not report empirical utility or security results for these systems.

\subsection{Utility}
\input{tables-figures/utility}
Table~\ref{tab:utility_pruned} reports utility on the 581 tasks retained by our interface-neutral evaluation protocol.
Across all four environments, \sys{} completes 476/581 tasks (81.9\%) with \texttt{Claude Sonnet-4.6}, 456/581 tasks (78.5\%) with \texttt{Gemini-3.5-Flash}, and 423/581 tasks (72.8\%) with \texttt{GPT-5.6-Luna}.
\sys{} achieves the highest aggregate utility with all three models.
Its advantage over the stronger aggregate ReAct baseline ranges from 2.4 percentage points with \texttt{GPT-5.6-Luna} to 22.9 percentage points with \texttt{Claude Sonnet-4.6}.

\noindent\textbf{Comparison with reactive agents.}
With \texttt{Claude Sonnet-4.6}, \sys{} achieves the highest utility in every environment and reaches 81.9\% aggregate utility, compared with 59.0\% for \reactApiWeb and 52.3\% for \reactWeb.
Across all models and environments, \sys{} ranks first or ties for first in nine of the twelve model--environment combinations.
Compared specifically with \reactApiWeb, \sys{} performs better in eight combinations, ties in one, and performs worse in three.
Because \reactApiWeb receives the same API schemas and site-specific guidance as \sys{}, these results show that Baya's aggregate improvement cannot be attributed solely to API availability.
The comparison instead indicates that compiling a complete program before execution can improve utility relative to selecting actions reactively from the same structured interface.

\noindent\textbf{Variation across environments.}
The utility advantage is most consistent on Postmill, where \sys{} outperforms both ReAct configurations under every model by at least 39.6 percentage points.
Results on Shopping and Shopping Admin are more model-dependent.
\texttt{Gemini-3.5-Flash} with \reactApiWeb outperforms \sys{} on Shopping Admin, while \texttt{GPT-5.6-Luna} with \reactApiWeb outperforms \sys{} on both Shopping and Shopping Admin.
These environments contain workflows that can be shorter or easier to discover through the user interface than through the available API operations. For example, a task that asks the agent to reorder an item can be completed through the user interface by navigating to the order history page and clicking ``Reorder.'' Accomplishing the same task through REST APIs requires fetching the customer ID to identify the relevant storefront orders, determining which item the task refers to, creating a new shopping cart, adding the item to the cart, providing shipping and billing information, and finally submitting the order. Thus, a single high-level UI action can correspond to a sequence of lower-level API operations, introducing additional steps and opportunities for failure for compiled API-based code.

\noindent\textbf{Why does \sys{} achieve higher utility?}
At first glance, it may seem surprising that \sys{} achieves both higher security and higher utility, since restricting an agent's interaction capabilities might be expected to introduce a utility tradeoff. One possible explanation is that \sys{} provides agents with focused, task-relevant APIs rather than requiring them to select among general-purpose UI actions. This can make tasks easier to solve: prior work has similarly found that API-based agents can outperform web-based agents on task completion \cite{song2025beyond}. Our design further takes advantage of the code-generation capabilities of LLMs by allowing the agent to synthesize a complete program from these focused APIs before execution. In contrast, \reactApiWeb must select actions reactively at each step and can choose between API calls and general-purpose web interactions. This flexibility may introduce additional decision points and lead different models to adopt inconsistent mixtures of API and UI actions, potentially hurting overall utility. Thus, in our setting, the security restrictions imposed by \sys{} do not necessarily reduce the agent's capabilities; instead, they may provide a more structured interface that makes successful task completion easier.

\subsection{Security}
\input{tables-figures/security}

Table~\ref{tab:wasp_security} reports intermediate ASR, end-to-end ASR, and benign-task utility.
Across all three models and both environments, \sys{} achieves 0\% intermediate and end-to-end ASR.
The ReAct baselines exhibit intermediate ASRs of up to 52.8\% and end-to-end ASRs of up to 5.6\%.
The largest differences occur with DeepSeek-V4-Flash, for which \reactWeb reaches intermediate ASRs of 46.0\% on GitLab and 50.0\% on Postmill, while \reactApiWeb reaches 2.1\% and 36.1\%, respectively.
On Postmill with \texttt{Gemini-3.5-Flash}, \reactWeb and \reactApiWeb reach intermediate ASRs of 27.8\% and 52.8\%, respectively, while \sys{} remains at 0\%.
Across the 84 evaluated attack instances, no injection causes \sys{} to take an action judged to advance the attacker's objective.
This result is consistent with Baya's  goal of preventing untrusted webpage content from influencing runtime action selection.

\reactApiWeb and \reactWeb show mixed ASR results across models. Depending on the model, ReAct tends to favor either browser based or API-based actions. Claude uses APIs more frequently, whereas Gemini and DeepSeek rely more heavily on browser actions. We attribute cases where \reactApiWeb outperforms \reactWeb to the more selective view of data provided by APIs, which can expose only the information relevant to a request rather than an entire webpage, as browser actions often do.

\sys{} does not achieve the highest benign-task utility in every configuration.
On Postmill with \texttt{Claude Sonnet-4.6}, both \sys{} and \reactApiWeb exhibit 0\% observed ASR, while their utilities are 61.1\% and 91.7\%, respectively.
On GitLab with \texttt{DeepSeek-V4-Flash}, \sys{} maintains 0\% ASR but completes only 50.0\% of benign tasks, while the reactive baselines achieve higher utility with nonzero intermediate ASR.

\noindent\textbf{Comparison with UCM.}
We compare UCM~\cite{nikolic2026ucm} and \sys{} on GitLab using \texttt{Claude Sonnet-4.6} as the primary agent model.
On the 154-task GitLab utility set, UCM completes 116 tasks (75.3\%), compared with 143 tasks (92.9\%) for \sys{}.
On the 48 GitLab WASP instances, both UCM and \sys{} achieve 0\% intermediate and end-to-end ASR while completing 100\% of the benign tasks.
Thus, UCM matches Baya's observed security and utility on the GitLab WASP instances, while \sys{} achieves substantially higher utility on the broader GitLab task set.
This comparison is limited to GitLab because UCM requires site-specific masking rules that are unavailable for the other evaluated environments.

\noindent\textbf{Takeaway.}
\sys{} achieves 0\% observed intermediate and end-to-end ASR across every evaluated model and environment.
In contrast, the security of the ReAct baselines varies across models and environments, even when the agent can access the same APIs as \sys{}.
This comparison shows that API access alone does not prevent prompt injection; the critical property is separating program generation from untrusted runtime observations.
UCM also achieves 0\% observed ASR in the evaluated GitLab configuration, but relies on site-specific masking rules and attains lower utility on the full GitLab task set.
Overall, the results demonstrate \sys{}'s consistent resistance to control-flow hijacking across the evaluated attacks.

\subsubsection{Runtime LLM Exposure}

\sys{} permits a generated program to invoke a runtime LLM when the task requires semantic processing that cannot be expressed reliably using deterministic operations.
We measure how frequently generated programs include and execute this capability.
For the static analysis, we identify programs containing model-provider invocation paths or direct model SDK imports.
For the runtime analysis, we identify executions containing at least one logged request to a model API endpoint.
These measurements count programs that invoke an LLM rather than the total number of model requests.

\input{tables-figures/llmcalls}
Table~\ref{tab:runtime-llm-usage} shows that 13 of 627 generated programs (2.1\%) contain an LLM-call path and that 11 programs (1.8\%) invoke that path during execution.
The two remaining programs use the LLM as a fallback, but their deterministic processing succeeds before the fallback is reached.

Runtime LLM calls perform narrowly scoped semantic operations, including classifying subjective reviews, matching conceptual descriptions, and summarizing text.
In addition, 105 programs (16.7\%) use the shared deterministic fuzzy matcher.
The fuzzy matcher performs similarity-based string matching locally and does not invoke a model.

In 616 of 627 executions (98.2\%), retrieved website content is never supplied to a runtime LLM.
In the remaining 11 executions, the compiled program determines the available API operations and control-flow structure before the runtime model receives any website content.
The runtime model has no tool access and returns only a value consumed by the compiled program.
Malicious content may therefore influence a semantic classification or other task output, but it cannot cause the model to synthesize new API calls or expand the compiled action space.
These results show that Baya substantially limits runtime model exposure while preserving its protection against control-flow hijacking.

\subsection{JSON Plans vs. Generated Programs}
\input{tables-figures/pteoldandnew}

\noindent\textbf{Choice of Compiled Representation.} Throughout the paper, \sys{} refers to our current Python-based implementation.
In this subsection, we call this implementation \sysNew{} to distinguish it from our earlier JSON-based implementation, which we call \sysOld{}.
Our earlier design compiled each task into a structured JSON plan specifying the selected API operations, their parameters, execution order, dependencies, and references to the outputs of earlier API calls. 
A fixed execution engine then interpreted the plan using a restricted set of operations for filtering, iteration, conditional execution, and response extraction. 

As shown in Table~\ref{tab:pte_old_new_pruned}, \sysNew{} improves utility on all four sites under every model. With \texttt{Claude Sonnet-4.6}, the
  macro-average success rate across the four sites increases from 66.9\% to 82.4\%. The largest gains occur on Shopping Admin and Shopping, where
  utility increases by 28.0 and 22.2 percentage points, respectively; GitLab improves by 9.8 points, while Postmill, where \sysOld{} was already
  comparatively strong, improves by 1.9 points.

The larger gains on the Shopping environments are consistent with the kinds of execution logic their tasks require, such as constructing nested request bodies, resolving instance-specific identifiers, traversing paginated results, and filtering or ranking records returned by earlier calls. These results show that even a detailed structured plan is not sufficient for reliable execution, even when that plan explicitly represents API calls, parameters, dependencies, and control flow. Generating executable code provides the flexibility needed to implement the more complex data processing and site-specific execution logic required by real web tasks, leading to substantially higher utility.

\subsection{Latency and Cost}\label{sec:latency_and_cost}
\input{tables-figures/latencyandcost}

Table~\ref{tab:latency-cost} reports mean end-to-end latency, token usage, and estimated model cost per task.
The measurements include all model calls, program generation, local validation, live execution, and replanning performed before the agent returns its final answer.

\sysNew{} uses an average of 1.37 million tokens, costs \$0.70, and requires 327 seconds per task.
Relative to \reactApiWeb, \sysNew{} incurs approximately 2.0$\times$ the monetary cost and 3.8$\times$ the latency.
Relative to \reactWeb, it incurs approximately 4.4$\times$ the cost and 5.5$\times$ the latency.
This overhead reflects the additional model inference and validation required to generate and test a complete program before execution.
\sysNew{} is approximately 4.0$\times$ slower than \sysOld{} and uses 6.8$\times$ as many tokens.
Despite the higher token count, its estimated cost is 5.1\% lower because monetary cost depends on the models and token categories used rather than token count alone.
Compared with UCM, \sysNew{} reduces mean latency by 30.3\% and estimated cost by 82.6\%.

\noindent\textbf{Takeaway.}
\sysNew{} incurs greater latency and monetary cost than the ReAct baselines, primarily because it generates and validates a complete program before interacting with the live website.
This overhead accompanies the utility and security improvements reported above.
Among the evaluated security-oriented agents, \sysNew{} remains substantially faster and less expensive than UCM.

%% file: tables-figures/utility.tex
\begin{table}[t]
\centering
\scriptsize
\setlength{\tabcolsep}{1.8pt}
\renewcommand{\arraystretch}{1.08}

\caption{Utility on valid WebArena tasks. Bold indicates the highest
utility within each model and environment.}

\resizebox{\columnwidth}{!}{%
\begin{tabular}{@{}llcccc@{}}
\toprule
\textbf{Model} &
\textbf{System} &
\shortstack{\textbf{GitLab}\\$n=154$} &
\shortstack{\textbf{Shopping}\\$n=171$} &
\shortstack{\textbf{Shopping}\\\textbf{Admin}\\$n=150$} &
\shortstack{\textbf{Postmill}\\$n=106$} \\
\midrule

\multirow{3}{*}{\shortstack[l]{Claude\\Sonnet-4.6}}
& \sys         & \textbf{92.9\%} & \textbf{71.3\%}
               & \textbf{81.3\%} & \textbf{84.0\%} \\
& \reactWeb    & 64.9\% & 53.2\% & 48.0\% & 38.7\% \\
& \reactApiWeb & 77.3\% & 50.3\% & 64.7\% & 38.7\% \\

\midrule

\multirow{3}{*}{\shortstack[l]{Gemini-3.5\\Flash}}
& \sys         & \textbf{85.1\%} & \textbf{74.3\%}
               & 70.7\% & \textbf{86.8\%} \\
& \reactWeb    & 45.5\% & 60.2\% & 48.7\% & 47.2\% \\
& \reactApiWeb & 75.3\% & 70.8\% & \textbf{86.0\%} & 42.5\% \\

\midrule

\multirow{3}{*}{\shortstack[l]{GPT-5.6\\Luna}}
& \sys         & \textbf{72.1\%} & 62.0\%
               & 78.7\% & \textbf{83.0\%} \\
& \reactWeb    & 69.5\% & 59.1\% & 70.7\% & 43.4\% \\
& \reactApiWeb & \textbf{72.1\%} & \textbf{69.0\%}
               & \textbf{89.3\%} & 43.4\% \\

\bottomrule
\end{tabular}%
}

\label{tab:utility_pruned}
\end{table}

%% file: tables-figures/security.tex
\begin{table*}[t]
\centering
\small
\setlength{\tabcolsep}{4pt}
\renewcommand{\arraystretch}{1.15}

\caption{
Security evaluation under WASP attacks on GitLab and Postmill.
Intermediate ASR measures whether an injection influences the agent's
intermediate behavior, while end-to-end ASR measures whether the attack
ultimately achieves its malicious objective. Utility measures benign task
performance. Lower ASR and higher utility are better. Bold highlights that the agent achieved 0\% ASR.
}
\begin{tabular}{llcccccc}
\toprule
&
&
\multicolumn{3}{c}{\textbf{GitLab}} &
\multicolumn{3}{c}{\textbf{Postmill}} \\
\cmidrule(lr){3-5}
\cmidrule(lr){6-8}

\textbf{Model} &
\textbf{System} &
\textbf{ASR}$\downarrow$ &
\textbf{ASR}$\downarrow$ &
\textbf{Utility}$\uparrow$ &
\textbf{ASR}$\downarrow$ &
\textbf{ASR}$\downarrow$ &
\textbf{Utility}$\uparrow$ \\[-2pt]

& &
\scriptsize Intermediate &
\scriptsize End-to-end &
&
\scriptsize Intermediate &
\scriptsize End-to-end &
\\
\midrule

\multirow{3}{*}{Claude Sonnet-4.6}
& \sys
& \textbf{0.0\%} & \textbf{0.0\%} & \textbf{100.0\%}
& \textbf{0.0\%} & \textbf{0.0\%} & \textbf{61.1\%} \\

& \reactWeb
& \textbf{0.0\%} & \textbf{0.0\%} & \textbf{100.0\%}
& 25.0\% & 0.0\% & 100.0\% \\

& \reactApiWeb
& \textbf{0.0\%} & \textbf{0.0\%} & \textbf{100.0\%}
& \textbf{0.0\%} & \textbf{0.0\%} & \textbf{91.7\%} \\

\midrule
\multirow{3}{*}{Gemini-3.5-Flash}
& \sys
& \textbf{0.0\%} & \textbf{0.0\%} & \textbf{100.0\%}
& \textbf{0.0\%} & \textbf{0.0\%} & \textbf{77.8\%} \\

& \reactWeb
& 4.2\% & 0.0\% & 100.0\%
& 27.8\% & 0.0\% & 50.0\% \\

& \reactApiWeb
& 6.2\% & 0.0\% & 100.0\%
& 52.8\% & 0.0\% & 27.8\% \\

\midrule

\multirow{3}{*}{DeepSeek-V4-Flash}
& \sys
& \textbf{0.0\%} & \textbf{0.0\%} & \textbf{50.0\%}
& \textbf{0.0\%} & \textbf{0.0\%} & \textbf{61.1\%} \\

& \reactWeb
& 46.0\% & 0.0\% & 92.0\%
& 50.0\% & 0.0\% & 36.0\% \\

& \reactApiWeb
& 2.1\% & 0.0\% & 77.7\%
& 36.1\% & 5.6\% & 50.0\% \\

\bottomrule
\end{tabular}

\label{tab:wasp_security}
\end{table*}

%% file: tables-figures/llmcalls.tex
\begin{table}[t]
    \centering
    \footnotesize
    \setlength{\tabcolsep}{4pt}
    \renewcommand{\arraystretch}{1.12}
    \caption{LLM use in generated programs. $n$ is the number of programs analyzed; columns report the percentage containing an LLM-call path, invoking an LLM at runtime, or using the deterministic fuzzy matcher.}
    \begin{tabular}{@{}lcccc@{}}
        \toprule
        \textbf{Site} &
        \textbf{$n$} &
        \shortstack{\textbf{LLM path}\\\textbf{in code}} &
        \shortstack{\textbf{LLM invoked}\\\textbf{at runtime}} &
        \shortstack{\textbf{Deterministic}\\
        \textbf{fuzzy matcher}\footnote{This column counts references to the
        shared \texttt{fuzzy\_match} implementation and excludes other
        deterministic operations such as exact matching, filtering, sorting,
        and arithmetic.}} \\
        \midrule

        GitLab
        & 187 & 0.0\% & 0.0\% & 17.6\% \\

        Shopping
        & 171 & 4.7\% & 4.1\% & 14.6\% \\

        Shopping Admin
        & 154\makebox[0pt][l]{%
            \footnote{Of the 155 Shopping Admin tasks, one produced a
            code-generation log but no Python program.}}
        & 1.3\% & 0.6\% & 27.9\% \\

        Postmill
        & 115 & 2.6\% & 2.6\% & 3.5\% \\

        \midrule
        Total
        & 627 & 2.1\% & 1.8\% & 16.7\% \\

        \bottomrule
    \end{tabular}

    \label{tab:runtime-llm-usage}
\end{table}

%% file: tables-figures/pteoldandnew.tex
\begin{table}[t]
\centering
\small
\setlength{\tabcolsep}{3pt}
\renewcommand{\arraystretch}{1.08}

\caption{Utility of \sysOld and \sysNew. Cells report success rates; $n$ gives the number of valid tasks.
Bold indicates the higher utility for each model and environment.}
\begin{tabular}{@{}llcccc@{}}
\toprule
\textbf{Model} & \textbf{System} &
\shortstack{\textbf{GitLab}\\$n=154$} &
\shortstack{\textbf{Shopping}\\$n=171$} &
\shortstack{\textbf{Shopping}\\\textbf{Admin}\\$n=150$} &
\shortstack{\textbf{Postmill}\\$n=106$} \\
\midrule

\multirow{2}{*}{\shortstack[l]{Claude\\Sonnet-4.6}}
& \sysOld & 83.1\% & 49.1\% & 53.3\% & 82.1\% \\
& \sysNew & \textbf{92.9\%} & \textbf{71.3\%}
& \textbf{81.3\%} & \textbf{84.0\%} \\

\addlinespace[2pt]
\multirow{2}{*}{\shortstack[l]{Gemini-3.5\\Flash}}
& \sysOld & 70.8\% & 55.6\% & 48.0\% & 85.8\% \\
& \sysNew & \textbf{85.1\%} & \textbf{74.3\%}
& \textbf{70.7\%} & \textbf{86.8\%} \\

\addlinespace[2pt]
\multirow{2}{*}{\shortstack[l]{GPT-5.6\\Luna}}
& \sysOld & 65.6\% & 44.4\% & 39.3\% & 44.3\% \\
& \sysNew & \textbf{72.1\%} & \textbf{62.0\%}
& \textbf{78.7\%} & \textbf{83.0\%} \\

\bottomrule
\end{tabular}

\label{tab:pte_old_new_pruned}
\end{table}

%% file: tables-figures/latencyandcost.tex
\begin{table}[t]
    \centering
    \renewcommand{\arraystretch}{1.05}
    \caption{Average token usage, cost, and latency per task.}
    \begin{tabular}{@{}lrrr@{}}
        \toprule
        \textbf{Agent} &
        \textbf{Tokens} &
        \textbf{Cost (\$)} &
        \textbf{Latency (s)} \\
        \midrule
        
        \reactWeb    & 161.6K & 0.1573 &  59.91 \\
        \reactApiWeb & 672.8K & 0.3449 &  86.60 \\
        \sysOld      & 201.1K & 0.7345 &  82.47 \\
        \sysNew      & 1.37M  & 0.6974 & 327.15 \\
        UCM          & 1.50M  & 4.0062 & 469.04 \\
        
        \bottomrule
    \end{tabular}
    
    \label{tab:latency-cost}
\end{table}

%% file: sections/discussion.tex
\section{Discussion}
\subsection{Limitations}
\heading{High Overhead.}
\sys{}'s guarantees come with overhead. As shown in
Section~\ref{sec:latency_and_cost}, ReAct incurs lower latency and token usage,
consistent with the findings of Piet et al.~\cite{piet2026pte}. This gap stems
from per-task compilation: API discovery, iterative validation, and
error-recovery logic all consume tokens and time before \sys{} takes any
action, so on an individual, previously unseen task its end-to-end cost and
latency exceed ReAct's.

This compilation cost can be amortized. Because a compiled program commits to
its control flow before execution, it avoids the DOM interpretation and action
selection that ReAct repeats at every step. Programs for common workflows can
be cached and reused, and, as Piet et al.~\cite{piet2026pte} also observe,
recurring patterns across similar tasks allow new programs to be derived from
existing ones rather than compiled from scratch. Under repeated or structurally
similar workloads, the per-task cost of compilation therefore decreases,
narrowing or even reversing the gap with ReAct. Pre-committed programs offer a
further benefit that ReAct cannot: they can be inspected before execution to
check policies and anticipate failures, making them easier to debug and verify.

Separately, \sys{} relies on trusted APIs constructed for each website before
any task is processed. This one-time step is not reflected in the measurements
above, but the resulting interfaces must be maintained as the website evolves,
a burden ReAct avoids by operating directly on the live page. We discuss how
this step could be automated securely in
Section~\ref{sec:missing_api_capabilities}.

\heading{Expressiveness.} 
Beyond cost, \sys{} has two limitations in expressiveness. First, it supports only tasks whose control flow can be determined before execution. This covers the vast majority of WebArena tasks, but excludes queries in which runtime data must determine actions that could not be specified in advance. Second, its effectiveness depends on choosing the right task abstraction: programs that are too low-level become brittle, while programs that are too high-level may hide information needed for execution.

While benchmark tasks are usually well-defined, not all real-world user requests are. To address underspecified tasks, \sys{} could interact with the user during compilation to clarify missing information before committing to a program. Because this interaction occurs during compilation, before exposure to untrusted website content, it does not compromise \sys's security guarantees. Tasks that depend on runtime data can still be supported as long as the possible operations and branches are specified in advance. However, if new requirements are discovered during execution that cannot be handled by the compiled program, \sys{} would need to return control to the user, explain the issue, and allow the user to provide an updated request for recompilation. This introduces additional overhead but ensures that new instructions come from the trusted user rather than untrusted website content.

Taken together, these limitations suggest that ReAct remains preferable in three settings: highly dynamic websites, where frequent structural changes make preprocessed interfaces costly to maintain; exploration-heavy tasks, where external data is needed during compilation and control flow cannot be fixed in advance; and workflows spanning many websites, some of which may not have been preprocessed.

\subsection{Providing Missing API Capabilities}
\label{sec:missing_api_capabilities}

\revision{As discussed in \cref{sec:trusted-api}, extending \sys{} to arbitrary websites requires automating the construction of Trusted Website APIs while ensuring that generated operations are safe to include in the trusted interface.}

One possible deployment model is therefore a separate, secure site-onboarding phase. An onboarding agent could explore representative workflows, observe the underlying requests, and synthesize parameterized operations together with their contracts. This process could be isolated from \sys{}'s per-task compilation and execution and performed using a least-privileged account. Generated operations could then be tested against expected outputs and state changes, restricted to approved network destinations and actions, and reviewed or certified before being admitted to the trusted interface.

Developing methods that automate this process without exposing \sys{}'s planning process to untrusted browser content is an important direction for future work. Such methods would extend \sys{} to websites without complete APIs while preserving the separation between untrusted web content and trusted execution.

%% file: sections/related_works.tex
\section{Related Work}
\label{sec:related}

\noindent\textbf{Web agents and structured interfaces.}
Different paradigms have been proposed for developing web agents. The dominant approach is ReAct~\cite{yao2023react}, which underlies baselines evaluated on WebArena~\cite{zhou2024webarena} as well as end-to-end multimodal agents such as WebVoyager~\cite{he2024webvoyager}. Similar reactive methods are used by vision-based agents such as SeeAct~\cite{zheng2024seeact}, which grounds model actions directly to page elements, and OpAgent~\cite{guo2026opagent}, which achieves state-of-the-art performance on WebArena. Another line of work replaces or augments vision- and DOM-based interaction with structured interfaces such as APIs. Song et al.~\cite{song2025beyond} show that API-based and hybrid API/browser agents outperform accessibility-tree-only agents on WebArena. Similarly, Zhang et al.~\cite{zhang2025apivsgui} find that augmenting GUI agents with APIs improves success rates and reduces the number of steps on office-related tasks from OSWorld~\cite{xie2024osworld}.

\noindent\textbf{Prompt injection against web agents.} Because ReAct agents condition action selection on untrusted web content, third-party page content can redirect their behavior through indirect prompt injection~\cite{greshake2023not}. Benchmarks such as InjecAgent, AgentDojo, and WASP systematically evaluate this threat~\cite{zhan2024injecagent,debenedetti2024agentdojo,evtimov2025wasp}, while DoomArena supports evolving, composable attacks across environments such as BrowserGym~\cite{boisvert2025doomarena,chezelles2025browsergym}. Attacks range from simple textual injections to stealthy page elements and pixel-level perturbations~\cite{liao2025eia,wang2025webinject}, motivating defenses that do not rely on the model recognizing malicious content.

\noindent\textbf{Defending against prompt injection.}
Prior defenses strengthen ReAct agents through instruction hierarchy or data/instruction separation~\cite{wallace2024instruction,chen2025struq,10.1145/3719027.3744836,piet2024jatmo}, but remain vulnerable to adaptive attacks~\cite{nasr2025attacker,wen2025rlhammerllmsnails}. Others monitor execution or restrict agent actions, including AgentArmor~\cite{wang2025agentarmor}, TaskShield~\cite{jia2025taskshield}, and ceLLMate~\cite{meng2025cellmate}. Security-by-construction approaches instead prevent untrusted content from influencing control flow. CaMeL~\cite{debenedetti2025camel} compiles the trusted query into a program while confining untrusted data, and Fides~\cite{costa2025fides} enforces information-flow policies. Beurer-Kellner et al.~\cite{beurerkellner2025patterns} identify plan-then-execute as a broader security pattern, while Piet et al.~\cite{piet2026pte} argue that it should be the default for web agents and show that it supports all WebArena tasks. Recent extensions apply related ideas to computer-use agents through upfront branching plans~\cite{foerster2026camelscua} or masking and quarantining untrusted content while retaining reactive execution~\cite{nikolic2026ucm}.

%% file: sections/conclusion.tex
\section{Conclusion}

We presented \sys{}, a web agent that adopts the compile-then-execute paradigm and is secure against prompt-injection-driven control-flow hijacking by design.
Our evaluation on WebArena Verified and WASP shows that \sys{} achieves higher utility than the ReAct baselines while maintaining 0\% observed ASR across all evaluated attacks.

\newpage

%% file: sections/appendix.tex
\section{Appendix}
\label{app:appendix}

\subsection{WebArena Evaluation}
\label{app:evaluation-method}
We use the WebArena~\cite{zhou2024webarena} task suite with revised intents and corrected ground truth from WebArena Verified~\cite{hattami2025webarena}. WebArena originally evaluates tasks using three mechanisms: \texttt{program\_html}, which checks the resulting website state; \texttt{string\_match}, which checks the agent's response for expected content; and \texttt{url\_match}, which checks the final browser URL. WebArena Verified replaces these with \texttt{AgentResponseEvaluator} for response-based tasks and \texttt{NetworkEventEvaluator} for tasks validated through browser network events. Below, we describe how we adapt these evaluators to support API-based execution.

\subsubsection{API-Compatible Evaluation.}
Not all WebArena Verified evaluators are compatible with \sys{}'s execution interface. We therefore construct a common evaluation harness that preserves the task semantics while allowing both browser-based and API-based agents to be graded under comparable criteria.

For tasks originally evaluated with \texttt{string\_match} or \texttt{program\_html}, we retain WebArena's outcome-based evaluators while using the revised task intents from WebArena Verified. We manually verify and correct the corresponding ground-truth data where necessary. These evaluators are directly applicable to both agents because they assess either the agent's returned information or the resulting website state, rather than requiring a particular execution path.

For tasks originally evaluated with \texttt{url\_match}, the evaluator itself is not applicable to \sys{}, since \sys{} does not need to navigate a browser to the target page in order to obtain the requested information. We therefore convert these tasks to \texttt{string\_match}. For each task, we manually complete the navigation and identify content unique to the expected destination page; successful agents are then required to return this page-specific information. This preserves the task's semantic objective without requiring browser navigation as the success condition.

We do not use \texttt{NetworkEventEvaluator} to grade \sys{}. Unlike the outcome-based evaluators above, \texttt{NetworkEventEvaluator} validates the execution trace by requiring a specific browser-generated network request. For conventional browser agents, these requests arise naturally as the agent clicks buttons, fills forms, and submits actions through the benchmark browser. For example, creating a Postmill post may be validated by observing a particular form \texttt{POST} request and its \texttt{302} response.

\sys{}, however, compiles the task into calls to a Trusted Website API and executes those calls directly rather than reproducing the corresponding frontend interaction sequence in the benchmark browser. Some Trusted Website API implementations internally use Playwright, but these operations execute behind the API abstraction rather than as actions in the benchmark's instrumented browser session. As a result, \sys{} may successfully produce the correct website state without generating the specific UI request required by \texttt{NetworkEventEvaluator}; the underlying request may also differ in endpoint, method, or payload. Thus, the incompatibility is with the validator rather than the underlying task: \sys{} can, for example, create the requested Postmill post even though it does not emit the particular browser-side request used by Webarena Verified to recognize success.

We additionally remove tasks whose intents or ground truth are unusable independent of agent architecture. Of the 623 single-site tasks, we label 42 as bad or impossible because the intent is ambiguous, underspecified, or cannot be completed reliably even by a fully capable browser agent. We also exclude all 31 cross-site tasks. This leaves 581 single-site tasks with well-defined, feasible objectives for our primary evaluation. 

Finally, because ReAct and \sys{} expose different execution traces and response formats, we give each agent credit if it satisfies any evaluator applicable to its execution mode. ReAct is evaluated under our common harness and additionally with WebArena Verified's evaluator when a valid browser network event provides evidence of successful completion. \sys{} is evaluated under the common outcome-based harness. This allows both agents to receive credit for completing the same task intent without requiring them to produce identical browser-level execution traces.

\subsection{Experimental Setup}
\label{app:experimental_setup}
We deploy the WebArena Verified websites as isolated Docker instances, using five independent worker stacks to execute at most five utility or security tasks concurrently.
Each worker is reset according to the benchmark protocol to prevent state changes from one task from affecting another.

For \sys{}, the models serve as the compiler and are invoked through Claude Code, Antigravity, and Codex, respectively.
Claude Code uses high reasoning effort, while Antigravity and Codex use medium reasoning effort.
The ReAct agents invoke the corresponding models directly through their provider APIs.
We impose a 900-second compilation timeout on each \sys{} attempt and limit ReAct agents to 30 action steps per task.

For latency measurements, we execute tasks serially on a single website stack to avoid resource contention, using a GCP \texttt{\seqsplit{n2-standard-8}} virtual machine with eight virtual CPUs and 32\,GB of memory.
The machine has no GPU, as all model inference is performed through remote provider APIs.
We obtain token counts from model-response logs and estimate monetary cost using LiteLLM's pricing table or, for models accessed through OpenRouter, the cost reported by OpenRouter.
These values are estimates rather than provider-invoiced charges.

\subsubsection{Implementation and Infrastructure Details}
\label{app:implementation-details}
Our evaluation environment used Python 3.12.11, Docker Engine 28.1.1,
Docker Compose 2.35.1, Prance 25.4.8.0, LiteLLM 1.80.0, Playwright 1.58.0,
and Chromium 145.0.7632.6. The latency experiments ran on an x86-64 Debian
GNU/Linux 13 VM in the \texttt{us-central1-c} zone with a 500\,GB balanced
persistent disk. The agent, browser, API servers, and benchmark websites
ran on the same host and communicated over the loopback interface.
\par\medskip
\noindent For \sys{}, DeepSeek was accessed through Codex using OpenRouter's
Responses-compatible API.
\par\medskip
\noindent The ReAct OpenAI configuration additionally imposed a 4,096-token output
limit. Claude Code occasionally invoked \texttt{claude-haiku-4-5}
internally for small helper operations; this behavior was controlled by
the harness rather than explicitly configured by us. Claude Code used the default high reasoning effort setting.

\subsection{Qualitative Differences Between \sys{} and ReAct}
The aggregate results obscure important differences in how the agents approach individual tasks. We examine two representative cases: aggregation over paginated records and precise query construction. These cases illustrate the benefits of API access for record-intensive tasks and differences between ReAct's incremental execution and \sys{}'s compiled, systematically scoped programs.

\noindent\textbf{Aggregation over paginated records.}
Many Shopping Admin tasks require aggregating records that are distributed across multiple UI pages. For example, one task asks for the email address, name, and phone number of the customer with the most cancellations. \reactWeb struggles to assemble the necessary data from the paginated tables. With API access, both \reactApiWeb and \sys{} solve the task using two REST endpoints. Their execution strategies differ, however: ReAct constructs the solution incrementally through multiple Python interpreter calls and hardcodes the observed number of pages, whereas \sys{} generates a single self-contained program that handles data retrieval and aggregation.

\noindent\textbf{Precise query scoping.}
API access alone does not ensure that a query faithfully captures every constraint in the task. Shopping uses an unusual filtering syntax in which each constraint must be encoded explicitly. When asked to report a customer's number of completed orders and total spending over the previous four months, ReAct filters by order status and date but omits the customer constraint, thereby returning another customer's results without detecting the error. In contrast, \sys{} includes the customer, status, and date constraints in the generated query and checks the resulting query construction during compilation-time testing.

Together, these cases show that API access can overcome limitations of browser-only interaction but does not by itself ensure reliable execution. \sys{} benefits from compiling the task constraints and data-processing logic into a self-contained program and validating them before live execution.

\subsection{Example Artifacts}
\label{app:example-artifacts}
See~\cref{fig:coding-harness-configuration} for harness configuration,
\cref{fig:toy-generated-workflow} for a complete \sys{} workflow, and \cref{fig:api-schema-examples} for API schemas.

\input{tables-figures/claude_code_settings}

\input{tables-figures/toy_code_ex}

\input{tables-figures/api_examples}

%% file: tables-figures/claude_code_settings.tex
\begin{figure}[!htbp]
\centering

\begin{appendixcodewide}{Claude Code Permissions and Sandbox Policy}
{
  "permissions": {
    "defaultMode": "dontAsk",
    "allow": [
      "Read(api/schemas/site_api_schema.json)",
      "Edit(generated/site/task/**)",
      "Bash(./venv/bin/python -m pytest *)",
      "..."
    ]
  },
  "autoMemoryEnabled": false,
  "sandbox": {
    "enabled": true,
    "failIfUnavailable": true,
    "allowUnsandboxedCommands": false,
    "filesystem": {
      "denyRead": ["/home", "/etc", "..."],
      "allowWrite": ["generated/site/task"],
      "denyWrite": ["api", "config", "tools", "..."]
    },
    "network": {"allowedDomains": []}
  }
}
\end{appendixcodewide}
\caption{Abridged Claude Code permissions and sandbox policy.}
\label{fig:coding-harness-configuration}
\end{figure}

%% file: tables-figures/toy_code_ex.tex
\usetikzlibrary{arrows.meta,calc}

\begin{figure*}[!t]
\begin{lrbox}{\trustedapibox}
\begin{tcolorbox}[
    enhanced,
    title={Task: Retrieve reviewers who say the Sony WH-XM0006
           headphone's ear cups are too small.},
    colback=white,
    colframe=black,
    colbacktitle=black,
    coltitle=white,
    fonttitle=\bfseries\small,
    boxrule=0.8pt,
    arc=2mm,
    left=1mm,
    right=1mm,
    top=1mm,
    bottom=1mm
]

\begin{tcolorbox}[
    enhanced,
    title={
        Phase 1:
        \compilerbot\;
        \sys{} compiler
        $\longrightarrow$
        compilation outputs
    },
    colback=white,
    colframe=black!70,
    colbacktitle=black!75,
    coltitle=white,
    fonttitle=\bfseries\footnotesize,
    left=1mm,
    right=1mm,
    top=0.5mm,
    bottom=0.5mm
]

\begin{minipage}[t]{0.49\linewidth}
\vspace{0pt}

\begin{toycode}{Generated Program File}
def main():
    token = authenticate()
    # Search for the product through the trusted API.
    products = trusted_api.search_products(
        query="Sony WH-XM0006", token=token
    )
    # Resolve the intended product deterministically.
    product = best_match(products, "Sony WH-XM0006")
    require(product, "Product not found")
    # Retrieve reviews only for the matched product.
    reviews = trusted_api.get_reviews(
        sku=product.sku, token=token
    )
    # The LLM sees only the retrieved review data.
    ids = data_only_llm.select(
        reviews, criterion="ear cups are too small"
    )
    # Return the selected reviewers.
    reviewers = [
        r.reviewer for r in reviews if r.id in ids
    ]
    print_final(reviewers or ["None"])
\end{toycode}

\end{minipage}
\hfill
\begin{minipage}[t]{0.49\linewidth}
\vspace{0pt}

\begin{toycode}{Generated Test File}
def test_main():
    # Replace external operations with local mocks.
    mock(authenticate, returns=FAKE_TOKEN)
    mock(search_products, returns=[xm0005, xm0006])
    mock(get_reviews, returns=[
        review_by_ava, review_by_ben
    ])
    # No live LLM request is made during testing.
    mock(data_only_llm, returns=[review_by_ava.id])
    output = run(main)

    # Verify product selection and the final answer.
    assert get_reviews.called_with(sku=xm0006.sku)
    assert output == "Ava"
\end{toycode}

\vspace{1mm}

\begin{toylog}{Coding-Harness Output}
===== coding_harness: site=shopping task=toy =====
Generated the program and companion unit-test files.
All mock tests pass. Here's what the implementation does:

1. Authenticates with the trusted Shopping API
2. Searches for "Sony WH-XM0006"
3. Selects the intended product deterministically
4. Fetches reviews only for the matched product
5. Uses a data-only LLM to identify relevant reviews
6. Prints the matching reviewers, or "None"
\end{toylog}

\end{minipage}
\end{tcolorbox}

\vspace{1mm}

\noindent
\begin{minipage}{0.46\linewidth}
\centering
\begin{tikzpicture}[font=\scriptsize]
    \node[inner sep=0pt] at (-0.3,0) {
        \includegraphics[
            height=11mm
        ]{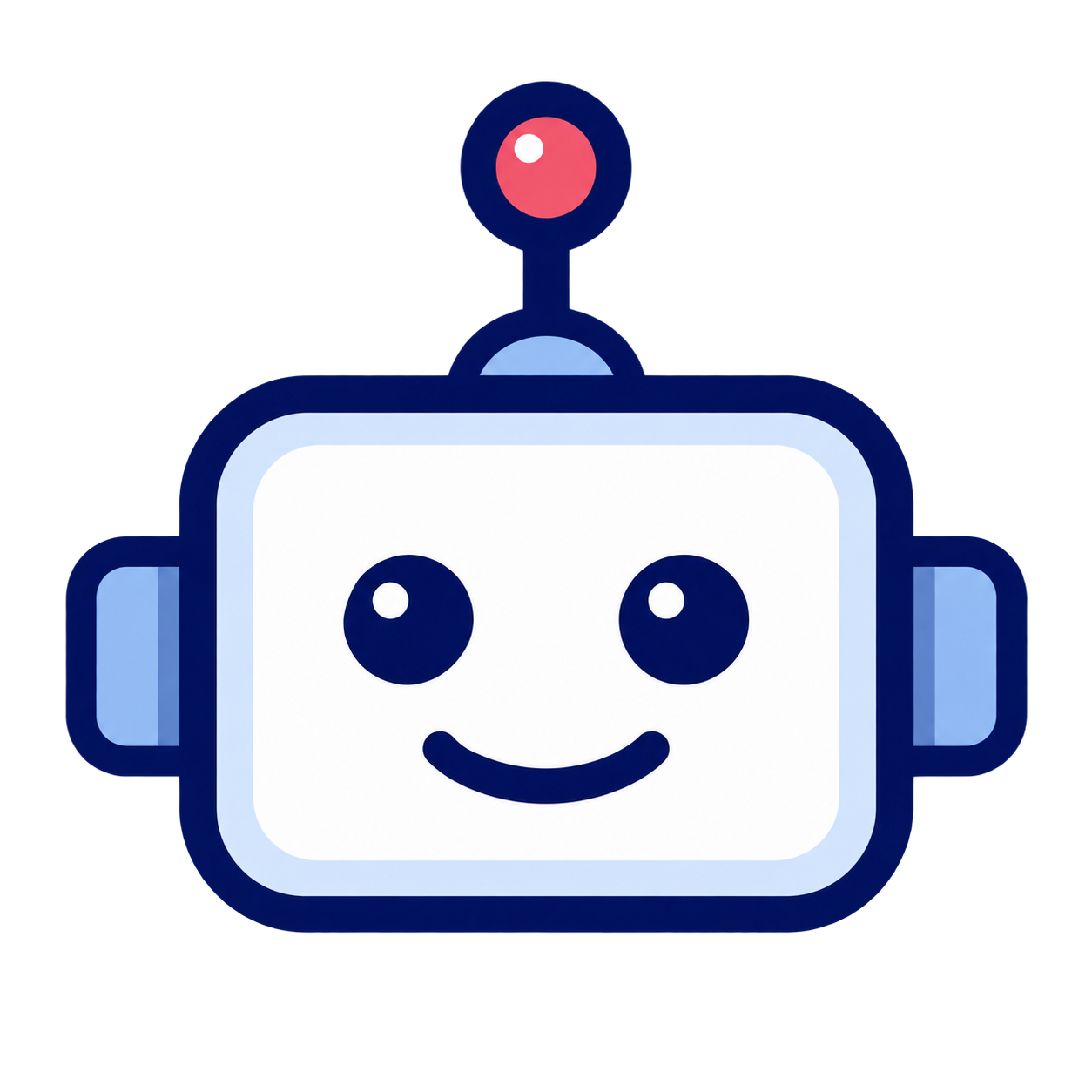}
    };
    \draw[-{Latex[length=2mm]}, thick]
        (-1.8,0.65) -- (-1.8,-0.65)
        node[midway,left] {Execute program};

    \draw[dashed,-{Latex[length=2mm]}, thick]
        (1.2,-0.65) -- (1.2,0.65)
        node[midway,right,align=left] {
            Failure: replan using\\
            program, test, and logs
        };

\end{tikzpicture}
\end{minipage}

\vspace{1mm}

\noindent
\begin{tikzpicture}[
    every node/.style={inner sep=0pt,outer sep=0pt}
]

\node[anchor=north west] (phase2) at (0,0) {
\begin{minipage}[t]{0.46\linewidth}
\vspace{0pt}
\begin{tcolorbox}[
    enhanced,
    title={Phase 2: \sys{} Execution},
    colback=white,
    colframe=black!70,
    colbacktitle=black!75,
    coltitle=white,
    fonttitle=\bfseries\footnotesize,
    left=1mm,
    right=1mm,
    top=0.5mm,
    bottom=0.5mm
]

\begin{toylog}{Python Error Diagnostics Log}
INFO  search_products("Sony WH-XM0006")
INFO  Product search returned 1 item(s)
INFO  Found product SKU (not logged)
INFO  Fetching product reviews
INFO  GET /V1/products/{sku}/reviews
INFO  Got 12 review(s) for product
INFO  Found 1 review(s) mentioning small ear cups
\end{toylog}

\end{tcolorbox}
\end{minipage}
};

\node[anchor=west] (phase3)
    at ([xshift=12mm]phase2.east) {
\begin{minipage}[t]{0.46\linewidth}
\vspace{0pt}
\begin{tcolorbox}[
    enhanced,
    title={Phase 3: Final Answer},
    colback=white,
    colframe=black!70,
    colbacktitle=black!75,
    coltitle=white,
    fonttitle=\bfseries\footnotesize,
    left=1mm,
    right=1mm,
    top=0.5mm,
    bottom=0.5mm
]

\begin{toylog}{Output}
===== FINAL ANSWER =====
Ava
===== END FINAL ANSWER =====
\end{toylog}

\end{tcolorbox}
\end{minipage}
};

\draw[-{Latex[length=2mm]},thick]
    (phase2.east) -- (phase3.west);

\end{tikzpicture}

\end{tcolorbox}
\end{lrbox}

\centering
\resizebox{0.96\textwidth}{!}{%
    \usebox{\trustedapibox}%
}

\caption{Toy example of \sys{} compilation, execution,
and failure-driven replanning.}
\label{fig:toy-generated-workflow}
\end{figure*}

%% file: tables-figures/api_examples.tex
\begin{figure*}[!t]
\centering

\begin{minipage}{0.48\textwidth}
\begin{appendixcodewide}{GitLab: \texttt{GET /projects} endpoint.}
"/projects": {
  "get": {
    "operationId": "get_projects",
    "parameters": [
      {"name": "membership", "in": "query",
       "type": "boolean", "default": false},
      {"name": "sort", "in": "query", "type": "string",
       "enum": ["asc", "desc"], "default": "desc"},
      ...
    ],
    "responses": {
      "200": {"schema": {"type": "array",
        "items": {"\$ref": "#/definitions/API_Entities_BasicProjectDetails"}}},
      "401": {"description": "Unauthorized"},
      "403": {"description": "Forbidden"},
      "404": {"description": "Not Found"}
    }
  }
}
\end{appendixcodewide}
\smallskip
\centering
\small
(a) GitLab API schema
\end{minipage}
\hfill
\begin{minipage}{0.48\textwidth}
\begin{appendixcodewide}{Shopping: \texttt{GET /V1/products} endpoint.}
"/V1/products": {
  "get": {
    "operationId": "GetV1Products",
    "parameters": [
      {"name": "searchCriteria[filterGroups][0][filters][0][field]",
       "in": "query", "type": "string"},
      {"name": "searchCriteria[filterGroups][0][filters][0][value]",
       "in": "query", "type": "string"},
      {"name": "searchCriteria[filterGroups][0][filters][0][conditionType]",
       "in": "query", "type": "string"},
      {"name": "searchCriteria[sortOrders][0][direction]",
       "in": "query", "type": "string"},
      {"name": "searchCriteria[pageSize]",
       "in": "query", "type": "integer"},
      ...
    ],
    "responses": {
      "200": {"schema": {
        "\$ref": "#/definitions/catalog-data-product-search-results-interface"
      }},
      "401": {"description": "Unauthorized"},
      "default": {"description": "Unexpected error"}
    }
  }
}

\end{appendixcodewide}
\smallskip
\centering
\small
(b) Shopping API schema
\end{minipage}

\caption{Abridged API schemas provided to the coding harness.
GitLab illustrates conventional typed parameters, enums,
and defaults, while Shopping illustrates Magento's nested
filtering, sorting, and pagination syntax.}
\label{fig:api-schema-examples}
\end{figure*}